\documentclass[useAMS,usenatbib]{mn2e}

\usepackage{CJKutf8}
\usepackage{amsmath,graphicx,longtable,hyperref}
\usepackage{natbib,threeparttable,rotating,placeins}
\usepackage{ulem,changepage}
\usepackage{xcolor}
\hypersetup{
    colorlinks,
    linkcolor={red!50!black},
    citecolor={blue!50!black},
    urlcolor={blue!80!black}
}

\newcommand{\doug}[1]{{
{#1}}}

\newcommand{\change}[1]{{
{#1}}} 
\newcommand{\changes}[1]{{
{#1}}} 

\newcommand{\psitau}{{ \Psi\left( \tau | \lambda \right) }}

\newcommand{\taumean}{{\langle \tau \rangle}}

\newcommand{\cream}{{\texttt{CREAM} }}
\newcommand{\javelin}{{\texttt{JAVELIN}}}

\newcommand{\Tx}{{T_{\rm irr}}}
\newcommand{\Lx}{{L_{\rm LP}}}
\newcommand{\Fx}{{F_{\rm LP}}}
\newcommand{\Hx}{{H_{\rm LP}}}
\newcommand{\epslp}{{\epsilon_{\rm LP}}}

\newcommand{\be}{\begin{equation}}
\newcommand{\ee}{\end{equation}}
\newcommand{\bea}{\begin{eqnarray}}
\newcommand{\eea}{\end{eqnarray}}

\title[ Rimmed and Rippled Discs to Explain AGN Continuum Lags]{Rimmed and Rippled Accretion Disc Models to Explain AGN Continuum Lags}

\author[D.A.Starkey et al.]{{D. A. Starkey$^{1,2}$\thanks{E-mail:
ds207@st-andrews.ac.uk (DAS);
jhuan192@ucsc.edu (JH);
jiamu$\_$huang@ucsb.edu (JH);
kdh1@st-andrews.ac.uk (KDH); 
lin@ucolick.org (DL)}, 
Jiamu Huang$^{3,4}$, 
Keith Horne$^{1}$, 
Douglas N. C. Lin$^{3, 5}$}
\\
$^{1}$SUPA School of Physics and Astronomy, St Andrews, KY16 9SS, Scotland, UK\\
$^{2}$Department of Astronomy, University of Illinois at Urbana-Champaign, Urbana, IL 61801, USA\\
$^{3}$Department of Astronomy and Astrophysics, University of California, Santa Cruz, CA 95064, USA\\
$^{4}$Department of Physics, University of California, Santa Barbara, CA 93106, USA\\
$^{5}$Institute for Advanced Studies, Tsinghua University, Beijing, 100086, China
}

\begin{document}
\date{Accepted 2022~November~25. Received 2022~November~24; in original form 2022~May~29.}

\pagerange{\pageref{firstpage}--\pageref{lastpage}} \pubyear{2022}

\maketitle

\label{firstpage}

\begin{abstract}
We propose a solution to the problem of accretion disc sizes in active galactic nuclei being larger
when measured by reverberation mapping
than predicted by theory.
Considering the {\doug{disc's exposed-surface}} thickness profile $H(r)$,
our solution invokes a steep rim or rippled structures
irradiated by the central lamp-post.
We model the continuum lags and the faint and bright disc spectral energy distribution (SED) in the best-studied case NGC~5548 (black hole mass $M_\bullet=7\times10^7\,M_\odot$, disc inclination $i=45^\circ$).
With the lamp-post off, the faint-disc SED fixes
a low accretion rate $\dot{M}\simeq0.0014~M_\odot\,{\rm yr}^{-1}$ and high prograde black hole spin
$a_\bullet\simeq0.93$, for which $r_{\rm in}=2\,G\,M_\bullet/c^2$ and $L_{\rm disc}=0.25\,\dot{M}\,c^2$.
The bright-disc SED then requires a lamp-post luminosity $\Lx\simeq5\,\dot{M}\,c^2/(1-A)$ for disc albedo $A$.
Reprocessing on the thin disc 
with $T\propto r^{-3/4}$ gives time lags
$\tau \propto \lambda^{4/3}$
but 3 times smaller than observed.
Introducing a steep $H(r)$ rim, or multiple crests, near $r\sim5$ light days,
reprocessing on the steep centre-facing slope 
increases temperatures from $\sim1500$~K to
 $\sim6000$~K, and this
 increases optical lags to match the lag data.
Most of the disc surface maintains
the cooler $T\propto r^{-3/4}$ profile that matches the SED.
The bright lamp-post may be powered by magnetic links tapping the black hole spin.
The steep rim occurs near the disc's dust sublimation radius as in the ``failed disc wind model for broad-line clouds''. Lens-Thirring torques 
aligning the disc and black hole spin
may also raise a warp and associated waves.  {\doug{In both scenarios, the small 
density scale height implied by the inferred value of $H(r)$ suggests possible marginal gravitational instability in the disc.}}
\end{abstract}

\begin{keywords}
accretion, accretion discs – galaxies: active – galaxies: Seyfert – quasars: supermassive black
holes - individual: NGC~5548.
\end{keywords}

\section{Introduction}
Active Galactic Nuclei (AGN) are fascinating objects of study not least because they are the brightest compact objects in the known universe, reaching luminosities up to $10^{13}L_\odot$ \citep{ba15}. These extreme luminosities are realised by gaseous material spiralling toward a central super-massive black hole (SMBH). The gas falls into an accretion disc whose differential rotation causes viscous torques to heat the gas, which loses gravitational potential energy and angular momentum allowing it to accrete onto the black hole \citep{lyndenbell1969, ss73}. The viscous forces and radiation transfer cause extreme temperatures of up to several million Kelvin \citep{AP} in the inner part of the accretion disc. These discs produce highly variable continuum emission that is correlated across UV, optical and infrared wavelengths \citep{ma10,sh14,ro15,ed15}. Variability amplitudes appear larger in the AGN of fainter Seyfert 1 galaxies than their brighter cousins, quasars \citep{ru18}. The exact physics of the accretion onto AGN is not fully understood, a major obstacle being that the compact sizes and remote distances of AGN accretion discs make them impossible to resolve with current or near-future telescopes. 

Nevertheless, spectral analysis is a useful tool to analyze the physical properties of AGN discs.  Viscous
dissipation in steady-state opaque regions of the disc generates an effective temperature with a distribution $T_{\rm e} \propto r^{-3/4}$ where $r$ is the 
disc radius. In these regions, the disc's spectral energy distribution is a composite of blackbody radiation from a range of radii and hence temperatures
\citep{pringle1981}. This distribution is indistinguishable from the reprocessed flux of a razor-thin passive disc around 
a bright central source.  Time-dependent evolution of either disc accretion rate or brightness of illuminating sources 
can lead to observational probes of the disc structure.

In this context,
reverberation mapping (RM; \citealt{bl82}) studies of continuum AGN light curves show that shorter-wavelength variations often lead those at longer wavelengths \citep{wa97,se05,ca07}. 
A linearised RM model predicts that the longer-wavelength continuum emission $F_\nu(\lambda, t)$ fluctuates about a mean level $\bar{F}_{\nu} \left( \lambda \right)$ and responds to the driving source $X(t)$ as
\begin{equation} 
\label{eqn:fnu}
	F_{\nu}(\lambda, t)= \bar{F}_{\nu} \left( \lambda \right)  
	+ \Delta F_{\nu}(\lambda)  
	\int_0^\infty \psitau \,
	X\left( t- \tau \right)\, d \tau
	\ .
\end{equation}
\change{Here $X(t)$ describes the shape of the driving lightcurve, 
and $\psitau$ is a distribution over $\tau$ that} describes a change in the accretion disc light observed at time $t$ in response to a change in the variable driving source at the  earlier time $t -\tau$. 
\change{
With $X(t)$ normalised to zero mean and unit variance, $\bar{F}_\nu(\lambda)$ is the mean and $\Delta F_\nu(\lambda)$ is the root-mean-square (RMS) of the responding lightcurve.} 
The lags $\tau$ arise due to light-travel-time delays and the full shape of the \change{delay distribution} 
encodes information on the accretion rate, disc inclination and slope of the disc radial temperature profile \citep{st17}. The dependence of the response function $\psitau$ on the accretion disc parameters allows us to probe the accretion disc structure despite direct resolution being currently impossible.

A simple accretion disc model invoked to describe continuum variability is known as the `lamp-post' model. Here, 
together with its intrinsic flux due to steady viscous dissipation and local emission,
a continuum variability arises from a thin accretion disc 
due to the thermal reprocessing of photons incident on the disc from a driving point source of irradiation, $X(t)$ 
in Eqn.~(\ref{eqn:fnu}), typically located several gravitational radii above the SMBH. The energy source of 
this `lamp-post' may be photons from the inner disc surface that are boosted to higher energies by Compton up-scattering
interactions with a corona of non-thermal electrons above the inner disc near the black hole.
This lamp-post model, combined with the $T\propto r^{-3/4}$ temperature profile of the standard thin disc,
gives rise to a $\tau \propto \lambda^{4/3}$ lag spectrum that may adequately explain continuum lags for a number of AGN \citep{ca07,st16,tr16}. However, other continuum variability studies suggest that this lamp-post model paints an incomplete picture of AGN variability. 

Recent continuum RM studies often broadly agree with the $\tau \propto \lambda^{4/3}$ picture, but find cross-correlation lags that are several (2--5) times larger than expected for a standard thin disc with independent measurements for black hole mass and Eddington ratio \citep{ca18,ho18,ed19}. Microlensing effects in multiply-imaged quasars similarly imply disc sizes larger than expected \citep{p08,m10,m13}.

This discrepancy between theory and observations is stimulating a search for helpful revisions to the standard thin-disc model, including two-stage reprocessing in a thick inner torus prior to disc irradiation
\citep{Gardner2017}, non-blackbody disc emission
due to a scattering atmosphere \citep{Hall2018},
replacing light travel time with slower alfven speeds \citep{Sun2020}, 
elevating the irradiating source several light days above the disc \citep{Kammoun2021} and 
diffuse bound-free and free-free continuum emission from the more extended emission-line regions \citep{Korista2001,Lawther2018,Chelouche2019,Netzer2022}.

These efforts have focused primarily on the overly-large disc sizes inferred from the measured lag spectrum $\tau(\lambda)$.
However, an equally relevant and in fact more easily measured constraint comes from the spectral energy distribution (SED) of the flux variations, $\Delta F_\nu(\lambda)$ in Eqn.~(\ref{eqn:fnu}), which provides an independent probe of the disc temperature profile.
In particular, a $T\propto r^{-\alpha}$ disc temperature profile corresponds not only to
an observable delay spectrum $\tau\propto \lambda^{1/\alpha}$, but also to
disc variations with a power-law SED, $\Delta F_\nu(\lambda)\propto\lambda^{(2-3\,\alpha)/\alpha}$.
For the standard model, $\alpha=3/4$, $\tau\propto\lambda^{4/3}$, and $\Delta F_\nu\propto\lambda^{-1/3}$.
A viable disc model needs to comply with both constraints, from the continuum lags, and from the SED.

In this work we explore one possible  interpretation for the large continuum lags observed in the STORM campaign on NGC~5548  \citep{fa16,st17} and subsequently in other targets \citep{ca18,ho18,ed19,hs20}. 
In particular, we show how lamp-post models with a finite disc thickness profile $H(r)$ can account for both the continuum lags $\tau(\lambda)$ and 
the variable disc SED $\Delta F_\nu(\lambda)$.
Section~\ref{sec:storm} summarises the constraints from the STORM campaign on NGC~5548, highlighting the strong challenge posed by these data to the standard thin-disc reprocessing model.
Section~\ref{sec:discs} reviews the lamp-post model for reprocessing from the surface of a finite-thickness accretion disc. 
A generalized vertical thickness profile $H(r)$ alters the accretion disc response function 
$\psitau$ by changing the light travel lag and the irradiation incidence angle on the reprocessing surface at each radius and azimuth.
Section~\ref{sec:sedfits} discusses 
thin-disc models for NGC~5548 that fit the disc SED in the faint and bright states, deriving the accretion rate in the faint state, the lamp-post luminosity in the bright state, and the disc inner radius, which constrains the black hole spin.
These razor-thin disc models, with $H(r)=0$,
fit the disc SED but predict lags far smaller than observed.
Considering power-law $H(r)$ profiles
in Section~\ref{sec:powerlawprofile},
and
rippled $H(r)$ profiles in Section~\ref{sec:ripples},
we find in both cases models that fit not only
the faint-disc and bright-disc SEDs, but also the inter-band continuum lags.
We summarise our findings and discuss some future prospects in Section~\ref{sec:conc}.

\section{Challenges from the STORM campaign on NGC~5548}
\label{sec:storm}

We focus our efforts on the strong challenges posed by the most complete RM study to date, based on data from
the Space Telescope and Optical Reverberation Mapping (STORM) campaign \citep{ro15,ed15,fa16,st17}. This experiment
realises very high signal-to-noise multi-wavelength multi-epoch light curves of the AGN within the Seyfert~1 galaxy NGC~5548.
The STORM campaign spanned 6 months in 2014, including X-ray, UV, and optical monitoring, achieving sub-day cadence in most bands.

\begin{figure}
\centering
\textbf{ \Large Probing the NGC~5548 disc}
\\
	\begin{tabular}{@{}c@{}}
 	\includegraphics[width=0.47\textwidth]{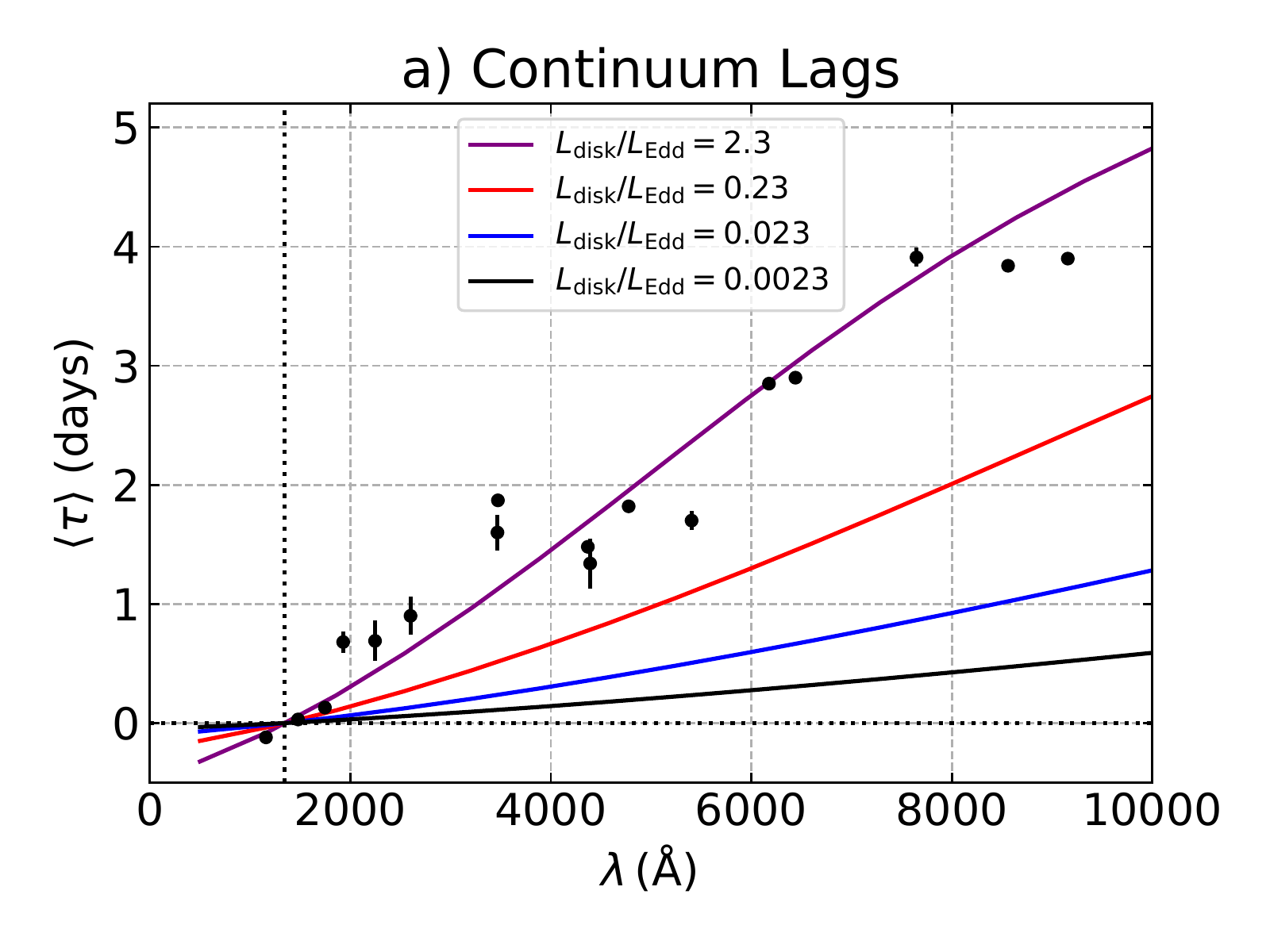}
\\
	\includegraphics[width=0.47\textwidth]{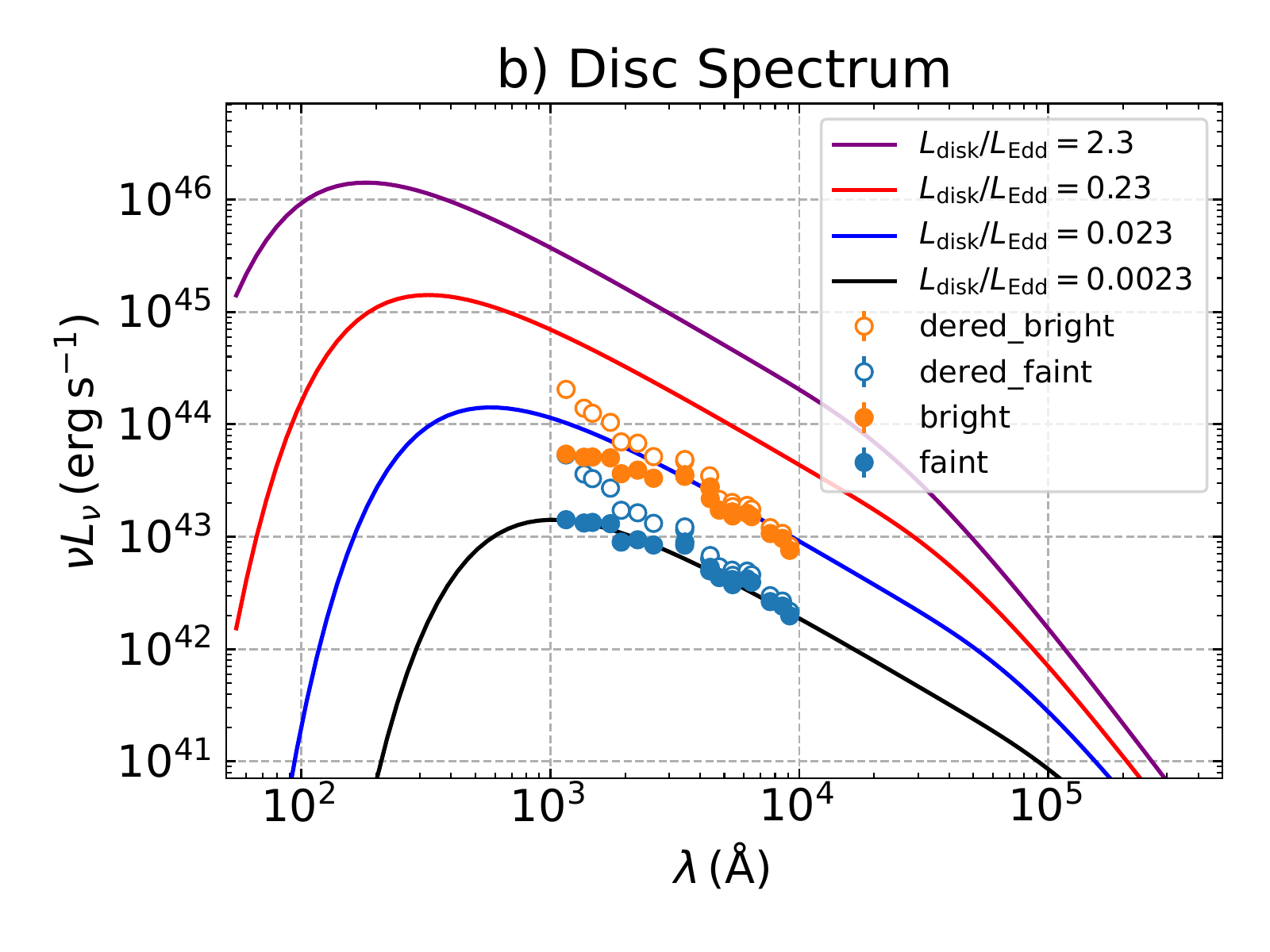}
	\end{tabular}
    \caption{
    Two independent probes of the NGC~5548 accretion disc derived from the STORM campaign, highlighting the failure of the standard disc model.
    \textbf{Panel~a) Lag spectrum}, $\tau(\lambda)$. The black data points with error bars show cross-correlation lags \citep{fa16}, measured relative to the HST 1367\AA\ light curve. 
    The model curves show lag spectra $\taumean(\lambda)$ for a razor-thin steady-state blackbody disc, a black hole mass
    $M_{\mathrm{BH}} = 7.0 \times 10^7\,M_\odot$ \citep{Horne2021},
    and Eddington ratios
    $L_{\rm bol}/L_{\rm Edd}=0.0023$, 0.023, 0.23 and 2.3.
 \textbf{Panel~b) Disc spectral energy distribution (SED)}, $\nu L_\nu(\lambda)$. 
 Filled circles show the observed disc spectra for NGC~5548 \citep{st17}
 at the faintest (blue) 
 and brightest (orange) states observed in the 2014 STORM campaign. Open circles show the data after de-reddening using SMC-like dust extinction with $E(B-V)=0.08$.
 The model curves give accretion disc SEDs for NGC~5548 ($D_{\rm L}=75$~Mpc) for $i=45^\circ$ \citep{Horne2021}.
 While the disc spectra are well fit by models with a low Eddington ratio, the disc lags require a much higher Eddington ratio.
}
\label{fig:storm}
\end{figure}

The multi-band continuum light curves, analysed using both cross-correlation techniques \citep{ed88,ed15} and the \javelin\ code \citep{zu11,fa16},
yield the basic lag spectra that we adopt for our analysis.
As shown in Fig.~\ref{fig:storm}a,
the measured lags, relative to the 1367\AA\ HST light curve, rise roughly linearly with $\lambda$, reaching
$\sim4$~days at 9000\AA.
Against this rising trend,
an excess lag of perhaps 0.5 days may be present in the $U$ band 
suggesting that Balmer continuum emission may contribute to the $U$-band lag. Our analysis ignores this feature and concentrates on fitting the overall rising trend.

In fitting the lag data, \cite{fa16} 
adopt a power-law model
\begin{equation}
 \tau = \tau_0 \, \left[ \left(\frac{\lambda}{\lambda_0}\right)^{4/3} - 1 \right]\ ,
\end{equation}
where $\lambda_0=1367$~\AA\ is the reference wavelength of the band used for the lag measurements, and compare the resulting 
parameter $\tau_0$ to the value predicted
for NGC~5548's black hole mass $M_\bullet = 5.2\times10^7\,M_\odot$
\citep{Pancoast2014}
and Eddington ratio $L_{\rm tot}/L_{\rm Edd}=0.1$. Here $L_{\rm tot}$
is the total (disc + lamp-post) luminosity
and $L_{\rm Edd}/L_\odot = 3.2 \times 10^4  M_\bullet/M_\odot$ is the Eddington luminosity, with $M_\odot$ and $L_\odot$ the
Sun's mass and luminosity.
In the \cite{fa16} analysis the best-fit lag parameter $\tau_0$, 
with a razor-thin disc model,
is larger than expected by a factor of $\sim3$
compared with the prediction for the thin-disc model.
The lags correspond instead
to a thin-disc model with $L_{\rm tot}/L_{\rm Edd}\sim3$.


\cite{st17} report a more stringent test of accretion disc theory using the \cream\ code \citep{st16} to fit the NGC~5548 STORM continuum light curves in detail.
\cream\
assumes a variable lamp-post irradiating the surface of a zero-thickness disc with 
an assumed power-law temperature profile, $T_{\rm e}(r)\propto r^{\alpha}$.
By fitting the model of Eqn.~(\ref{eqn:fnu}) to the light curve data at all wavelengths,
\cream\ determines
the light curve of the lamp-post variations $X(t)$,
the power-law temperature profile $T_{\rm e}(r)\propto r^\alpha$, the disc's inclination $i$,
and the mean and rms spectra, $\bar{F}_\nu(\lambda)$ and $\Delta F_\nu(\lambda)$, respectively.
This method senses not just the mean delay, but also the width and shape of the delay distribution $\psitau$, thus determining the disc inclination, $i=36^\circ\pm10^\circ$,
and the power-law index of the temperature profile, $\alpha = -0.99\pm0.03$.
The \cream\ analysis achieves a very satisfactory fit to the observed 
variations from $1150$~\AA\ to $9000$~\AA\ \citep{st17}.
This fit validates the adequacy of the linearised echo model, Eqn.~(\ref{eqn:fnu}), to describe
quite large observed disc variations, a factor 5 in brightness with little change in the shape of the SED.
However, the best-fit model's temperature profile is hotter and steeper than the $T_{\rm e}\propto (M_\bullet\,\dot{M})^{1/4}\,r^{-3/4}$ prediction of thin-disc theory,
and the observed delays are a factor of $\sim 3$ larger than that expected from a thin-disc
model with $L_{\rm tot}/L_{\rm Edd}\approx0.1$.
As a consequence, when the inferred $T_{\rm e}(r)$ profile is used 
to compute the disc SED, the 
model result is well above the observed SED.

\begin{table}
\begin{center}
\caption{NGC~5548 faint-state disc parameters
\label{tab:faintdisc}}
\begin{tabular}{|l|r| r| l}
\hline
Parameter & Previous value$^1$ 
& This work & Units
\\
\hline \hline
$M_\bullet$ & $5.2\times10^7$ & $7\times10^7$ & $M_\odot$
\\
$\dot{M}$ & $0.069$ & {0.0014} & $M_\odot\,{\rm yr}^{-1}$
\\ $a_\bullet$ & 0.0 & 0.93 &
\\
$r_{\rm in}$ & 6 & {2} & $r_\bullet$
\\
$\epsilon_{\rm disc}$ & $0.1$ & {0.25}
\\
$L_{\rm disc}$ & $0.1$ & {0.0023} & $L_{\rm Edd}$
\\
$i$  & $0^\circ$ & $45^\circ$
\\
$D_{\rm L}$ 
& 75 & 75 & Mpc
\\ 
\hline
\end{tabular}
\\$^1$ Previous values from \cite{fa16}.
\\
$L_{\rm Edd}=2.3\times10^{12}~(M_\bullet/7\times10^7M_\odot)\,L_\odot$.
\\
$r_\bullet/c=4.0\times10^{-3}~(M_\bullet/7\times10^7M_\odot)$ days.
\end{center}
\end{table}

To illustrate this problem,
Fig.~\ref{fig:storm} shows how the standard accretion disc model fails to simultaneously fit the observed disc lags and SED.
Fig.~\ref{fig:storm}a compares
the observed lags \citep{fa16}
with lags 
for standard thin-disc models at four accretion rates spaced by factors of 10.
Fig.~\ref{fig:storm}b compares these same
models with the
observed disc SED (host galaxy light subtracted) at the faintest and brightest flux levels observed during the STORM campaign \citep{st17}.
The models that fit the lags have accretion rates 100--1000 times higher than those that fit the bright-disc and faint-disc SEDs.

The observed faint-disc and bright-disc SEDs in Fig.~\ref{fig:storm} are corrected for Milky Way dust extinction, and are shown both with and without a correction for possible SMC-like dust in the host galaxy.
For this we adopt
$E(B-V)=0.08$, which raises the UV fluxes 
just enough to straighten the observed disc SED close to a power law. This indicates a range
of possible intrinsic disc SEDs and highlights that the red end near $10^4$~\AA\ is less affected than the UV end near $10^3$~\AA.
The observed disc SED is close to or slightly bluer than $\nu\,L_\nu \propto
\lambda^{-4/3}$ predicted by the standard disc theory.
Here $L_\nu= 4 \, \pi \, D_{\rm L}^2\, F_\nu$, where
$F_\nu$ is the observed flux density
and $D_{\rm L}=75$~Mpc is the luminosity distance to NGC~5548, based on its
redshift $z=0.0172$ \citep{ro15},
and a standard $\Lambda$CDM cosmology with 
$\Omega_{\rm M}=0.3$, 
$\Omega_\Lambda=0.7$ and 
$H_0 = 70\,\mathrm{km\,s^{-1}Mpc^{-1}}$.

For our analysis of NGC~5548, we adopt updated parameters
as summarised in Table~\ref{tab:faintdisc}, including
a somewhat larger mass, $7\times10^7M_\odot$,
and a disc inclination $i=45^\circ$,
as derived from emission-line velocity-delay maps \citep{Horne2021}.
The disc models shown in Fig.~\ref{fig:storm} are for a zero-thickness disc heated entirely by the viscous dissipation associated with accretion.
The model fitted to the faint-disc SED (black curve in Fig.~\ref{fig:storm}b) is discussed in Section~{\ref{sec:faintfit}.
In brief, the accretion rate $\dot{M}$ is adjusted to fit the red end of the SED, and the inner
disc radius $r_{\rm in}$ is adjusted to fit the UV end, where the flux decreases due to the maximum temperature achieved in the inner disc.

Having established this baseline model for the faint-disc SED, increasing $\dot{M}$ by a factor of 10
brings the model SED into agreement with the bright-disc SED.
This suggests that a 10-fold change in accretion rate, or corresponding increase in heating from the lamp-post irradiation, is needed to match the observed span of flux variations.
The lag data require a much higher accretion rate, and that
model is far above the observed disc SED.
This huge discrepancy
highlights the standard thin-disc model's failure to fit both the lags and the SED, and the importance of attending to both constraints rather than just the lags.

Our results in Fig.~\ref{fig:storm}
confirm the findings of \citep{fa16} and \citep{st17}.
It thus appears 
that the brightness temperature of the disc surface, which determines its surface brightness and thus the flux we observe, is lower 
than its colour temperature, which determines the wavelength of the reprocessed radiation at which we observe a given lag.
A low accretion rate that fits the SED predicts
lags smaller than observed (Fig.~\ref{fig:storm}a). A high accretion rate that fits the lags predicts a much brighter disc than observed (Fig.~\ref{fig:storm}b).
We address these seemingly incompatible constraints
by introducing a finite disc thickness, $H(r)$, and adjusting its shape to increase temperatures while reducing the fraction of the disc surface that is exposed to the lamp-post irradiation.
We thus turn to developing the framework of these finite-thickness disc models.


\section{Reverberating Accretion Disc Models}
\label{sec:discs}

We consider geometrically-thin accretion disc models in which differential rotation produces local viscous heat dissipation 
and outward transport of angular momentum that allows material to spiral inward and accrete onto the black hole. 
We neglect any auxiliary energy sources including those from embedded stars,
neutron stars, or stellar massive black holes \citep{artymowicz1993, lin1994, goodman2004} but we do take into account
the effect of surface irradiation, analogous to the context of protostellar discs \citep{chiang1997, garaud2007}.
A compact variable source of irradiation, modelled as a point source 
isotropically-emitting `lamp-post' at height $\Hx$ on the symmetry axis just above (and below)  the black hole, 
drives temperature variations that move outward through the disc to produce observable variability in the accretion disc 
SED. 
Possible contributors to such a  lamp-post are the X-ray and EUV photons reprocessed by the hot  electrons in the
corona above the disc close to the supermassive black hole.

\subsection{Steady disc irradiated by a lamp-post}

In a steady-state geometrically-thin accretion disc \citep{ss73},
the disc's effective temperature profile
$T_{\rm e}(r)$, found by balancing the viscous and irradiative heating
with blackbody radiation from both sides of the disc surface, can be written  
as 
\begin{equation}
\label{eq_trprof}
	T_{\rm e}^4 \left( r \right) 
	= \left( \frac{G\,M_\bullet \,\dot{M}}{2\,\pi\, \sigma \,r^3 }\right)
	 \, \left[ \frac{3}{4} \left( 1 - \sqrt{ \frac{ r_{\rm in}}{ r}} \right)
	 + \frac{\epslp \, r}{r_\bullet} f(r) \right]
	 \ .
\end{equation}
Here the left and right terms give
$T_{\rm v}^4(r)$ and $\Tx^4(r)$,
representing respectively contributions from viscous
heating and lamp-post irradiation.
In the pre-factor,
$G$ and $\sigma$ are the Newton and Stefan-Boltzmann constants,
$M_\bullet$ 
and $\dot{M}$ are the black hole's mass and accretion rate.

In the viscous heating term,
the factor $(1-\sqrt{r_{\rm in}/r})$ reduces viscous heating to zero at the inner radius $r_{\rm in}$ where 
the shear vanishes
(but see \cite{AgolKrolik2000}
for modification of this term including viscous torques due to magnetic links with the black hole).
For a black hole accretion disc the inner radius $r_{\rm in}$ is usually taken to be the inner-most stable circular orbit (ISCO) radius,
$r_{\rm in} = r_{\rm ISCO}$.  
\doug{ 
In terms of the black hole's angular momentum $J_\bullet$ and spin parameter $a_\bullet \equiv J_\bullet \,c/ G \, M_\bullet^2$,
\begin{equation}
    {r_{\rm ISCO} /  r_\bullet} = 3+Z_2 \mp [\,(3-Z_1) \,(3+Z_1 +2 \,Z_2) \,]^{1/2} 
    \ ,
\label{eq:riscoa}
\end{equation}
where $r_\bullet \equiv G\,M_\bullet / c^2$ is the gravitational radius,
with $c$ the speed of light,
$Z_1= 1 + (1-a_\bullet^2)^{1/3} [ (1+a_\bullet)^{1/3} + (1-a_\bullet)^{1/3}]$ and
$Z_2 = (3\, a_\bullet^2 +Z_1^2 )^{1/2}$ \citep{bardeen1972}. For black holes with
spin axis parallel/anti-parallel to the disc's angular momentum vector, the sign 
in Eqn.~(\ref{eq:riscoa}) is negative and positive  respectively.
For non-rotating black holes, $a_\bullet=0$, $Z_1=Z_2=3$, and $r_{\rm ISCO}=6 \,r_\bullet$.
For discs around rapidly spinning black holes with nearly parallel angular momentum vector, $a_\bullet 
\equiv 1 - \beta_{\rm a}$, $Z_1 \simeq 1 + (4\, \beta_{ \rm a})^{1/3} $, 
$Z_2 \simeq 2 +(\beta_{\rm a}/2)^{1/3}$, and  ${r_{\rm ISCO} /  r_\bullet} \simeq 1 + 3 (\beta_{\rm a}/2)^{1/3}$
so that $r_{\rm ISCO} / r_\bullet =2$ corresponds to $\beta_{\rm a} \simeq 2/27$ and $a_\bullet \simeq 0.93$
(Table \ref{tab:faintdisc}).}

In the irradiative heating term,
$\epslp$ is a dimensionless efficiency factor, and $f(r)$ describes the radial dependence of the lamp-post irradiation, as discussed below.
The dimensionless energy conversion and reprocessing efficiency, for the lamp-post model, is
\begin{equation}
	\epslp \equiv \frac{ \Lx \,(1-A)}{ \dot{M}\, c^2}
	\ ,
\end{equation}
where 
$\Lx$ is the bolometric luminosity of the lamp-posts above and below the black hole that irradiate the two sides of the disc, 
and $0<A<1$ is the albedo of the disc surface, assumed independent of $r$.
The lamp-post efficiency
$\epslp$ can differ from that of the disc,
\begin{equation}
    \epsilon_{\rm disc}
    \equiv \frac{ L_{\rm disc} } { \dot{M} \, c^2 } \ ,
\end{equation}
where, by integrating over both sides
its surface, the accretion disc's
bolometric luminosity is
\begin{equation}
L_{\rm disc} 
 = \int_{r_{\rm in}}^{r_{\rm out}}
  \sigma \, T_{\rm e}^4(r) \,
 4 \, \pi \, r \, dr \,
= \epsilon_{\rm disc} \, \dot{M}\, c^2
\end{equation}
With the lamp-post off ($\epslp=0$),
viscous heating alone gives $\epsilon_{\rm disc}=(r_\bullet/2\, r_{\rm in})$.
If we assume $r_{\rm in}= r_{\rm ISCO}$, then $\epsilon_{\rm disc}=1/12$ for a Schwarzshild black hole.
But, it is likely to be modified by radiative hydrodynamics of the disc gas \citep{jiang2014, jiang2019}.
The total luminosity,
radiated by both sides of the disc and by the lamp-posts above and below the disc, is
$L_{\rm tot} = L_{\rm disc} + \Lx$.

In the lamp-post model, the central source that irradiates the upper surface of the disc is idealised as an isotropic point source located on the symmetry axis.  \doug{ Based on constraints from x-ray reverbration-mapping and microlensing observations \citep{reis2013, cackett2021},
we consider models with} a height $\Hx \sim$ a few $(< 20)\,r_\bullet$} above the black hole.
The covering 
fraction $f(r)$ in Eqn.~(\ref{eq_trprof}) is then given, for $r\gg\Hx\sim r_{\rm in}$, by
\begin{equation}
	f(r)= r\, \frac{ \partial }{ \partial r} \left( \frac{ H(r) - \Hx }{ r } \right)
	\ .
\label{eq_covfrac}
\end{equation}
 $f(r)$ is a function of the assumed {\doug{ thickness profile $H(r)$ of the disc's 
 exposed surface}}, of which several are explored in this work.

The accretion disc SED is calculated using the Planck formula integrated over solid angle,
\begin{equation}
\label{eq_planck}
	F_\nu( \lambda)  
	= \int_{r_{\rm in}}^{r_{\rm out}}
	\frac{ 2\,\pi\, r \, dr \, \cos{i} } { D_{\rm L}^2}
	\frac{   
	2\,h\,c/\lambda^3}
	{\left[\, \exp{ \left( \frac{\displaystyle h\,c}{\displaystyle \lambda\,k_B\,T_{\rm e}(r)} \right)}- 1 \,\right]}
	\ ,
\end{equation}
where $\nu=c/\lambda$ is the photon frequency, $D_{\rm L}$ is the luminosity distance, $i$ is the disc inclination, 
$h$, $k_B$ and $c$ are the Planck constant, Boltzmann constant and speed of light, respectively.
This is for a flat disc, for which the foreshortening factor is $\cos{i}$, independent of position on the disc surface.
Our models for curved disc surfaces account for the foreshortening factor varying over the disc surface.
As is customary, we present observed and model disc SEDs in
\change{isotropic} luminosity units, 
\be
    \lambda\, L_\lambda(\lambda)
    =     \nu\,L_\nu(\lambda) 
    \equiv 4 \, \pi \, D_{\rm L}^2 \,\nu \, F_\nu( \lambda)
    \ .
\ee

\subsection{Continuum reverberation lags}

\label{sec:lags}

A sudden spike in irradiation from the central lamp-post source causes a lagged reverberation response from the disc, $\psitau$,
whose strength
and distribution in wavelength $\lambda$ and lag $\tau$ depends on several factors including the distance from the lamp-post to the disc surface element, the temperature profile $T_{\rm e}(r)$,
the lamp-post efficiency $\epslp$,
and the incidence angle of the lamp-post irradiation. To compute the total accretion disc response at a particular wavelength, we calculate the response at each disc surface element, and numerically integrate over the disc surface.

Observing the accretion disc at inclination angle $i$, the delay $\tau_{\rm lag}$
between observed photons emitted from the lamp-post 
and those reprocessed on the disc surface, at radius $r$, azimuth $\theta$ and height $H(r)$, is given by
\begin{equation}
\label{eq_taudel}
c \, \tau_{\rm lag} \left( r, \theta, i \right) = \sqrt{ \Delta h^2 + r^2} + \Delta h\, \cos{i} - r\, \cos{\theta}\, \sin{i}
\ , 
\end{equation}
where $\Delta h \equiv \Hx - H(r)$.
\change{This assumes Euclidean geometry and neglects general relativity effects near the black hole since $r_\bullet/c = 350$~s is small compared with the observed lags.}
Given Eqn.~(\ref{eq_taudel}) for the time delay and Eqn.~(\ref{eq_trprof})
for the temperature profile, 
our linearised reprocessing model for the continuum response 
at time delay $\tau$ and wavelength $\lambda$ is (see also \citealt{st17})

\begin{equation}
\label{eqpsitaulam}
	 \Psi_\nu (\tau | \lambda ) 
	 = \int d \Omega \, \frac{\partial B_{\nu} (T_{\rm e}, \lambda ) }{\partial T_{\rm e}} 
	 \, \frac{\partial T_{\rm e}}{\partial \Lx }
	 \, \frac{\partial \Lx }{\partial \Fx }
	 \, \delta \left( \tau - \tau_{\rm lag} (r, \theta, i ) \right)
 \ , 
\end{equation}

\noindent where $d\Omega$ is the solid angle on the observer's sky of a disc surface element
at radius $r$ and azimuth $\theta$,
$B_{\nu} (T_{\rm e}, \lambda )$ is the Planck function at the local effective temperature $T_{\rm e}(r)$, 
$\Lx$ is the lamp-post luminosity, $\Fx$ is the observed flux from the lamp-post, 
or in practice a proxy thereof,
$\delta(\cdot)$ is the Dirac delta function, and
$\tau_{\rm lag} (r,\theta,i)$ is the time delay given by Eqn.~(\ref{eq_taudel}). Observed inter-band continuum lags, obtained by cross-correlating light curves \citep{fa16}, correspond (approximately) to the response-weighted mean delay given by
\begin{equation}
\label{eq_taumean}
	\taumean \left( \lambda \right)
	\equiv \frac{\int_0^\infty \psitau \,\tau\, d \tau } { \int_0^\infty \psitau\, d \tau }.
\end{equation}

As discussed in \cite{st16}, for a flat disc geometry and $T_{\rm e}^4\propto M\, \dot{M} / R^3$, the mean delay scales as $\taumean \propto(M\,\dot{M})^{1/3}\,\lambda^{4/3}$
and is independent of the inclination $i$ even though $\psitau$ becomes wider and more highly skewed as the inclination increases from face-on ($i=0$) to edge-on ($i=90^\circ$) orientations.
For a concave bowl-like disc surface, however, the mean delay increases with inclination, as discussed in Section~\ref{sec:bowl} below.

These continuum lags (Eqn.~\ref{eq_taumean}) are set by the response function (Eqn.~\ref{eqpsitaulam}) which fundamentally depends on:

\noindent \hspace{1mm}
(i) the strength of the response at each disc surface element. This in turn depends on the irradiation and viscous heating effects resulting 
from the disc properties and geometry, and

\noindent \hspace{1mm}
(ii) the time lag (Eqn.~\ref{eq_taudel}) between the lamp-post irradiation and the reverberation response from each surface element on the disc. 
This lag depends on the accretion disc geometry.

\section{Zero-thickness disc models that fit the faint and bright disc SEDs}
\label{sec:sedfits}

The classic thin-disc model assumes a negligible vertical 
thickness $H(r) \simeq 0$ or $H(r) \ll \Hx$,
resulting in a $T_{\rm e}(r)$ profile given by
Eqn.~(\ref{eq_trprof}) with covering factor 
\begin{equation}
\label{eq_covflat}
	f(r) = \Hx/r
	\ .
\end{equation}
Since the entire disc surface is exposed to the variable lamp-post in
this model, $T_{\rm e}\propto r^{-3/4}$.  This $T_{\rm e} (r)$ distribution 
is similar to that predominately  due to viscous dissipation only,
$\epslp \ll 1$ in Eqn.~(\ref{eq_trprof}), which we assume holds during the faint state.  After the lamp-post is turned on, the 
elevated flux propagates outwards on light travel timescales $r/c$ much shorter than
the local dynamical time scale $
(r^3/G\,M_\bullet)^{1/2}$ at the radii $r$ reached by the 
light.  Consequently,  $H(r)$ cannot change significantly despite the intensified
surface heating.  Under these circumstances,
the irradiation increases the local effective temperature, and the resulting
changes in the continuum light curve then
have a wavelength-dependent lag time 
$\tau \propto 
\lambda^{4/3}$.

We adopt a 3-step approach in fitting the reverberating disc model to the NGC~5548 data. We first adjust $\dot{M}$ (Section~\ref{sec:faintfit})
and $r_{\rm in}$ (Section~\ref{sec:spin})
to fit the faint-disc SED with the lamp-post off. Then we turn on the lamp to fit the bright disc SED (Section~\ref{sec:brightfit}). Finally we 
adjust the $H(r)$ profile to fit the continuum lags (Sections~\ref{sec:powerlawprofile} and \ref{sec:ripples}).

 \begin{figure}
\centering
\textbf{ \Large Effect of $r_{\rm in}$ on the faint-disc SED}
\\
	\begin{tabular}{@{}c@{}}
 	\includegraphics[width=0.47\textwidth]{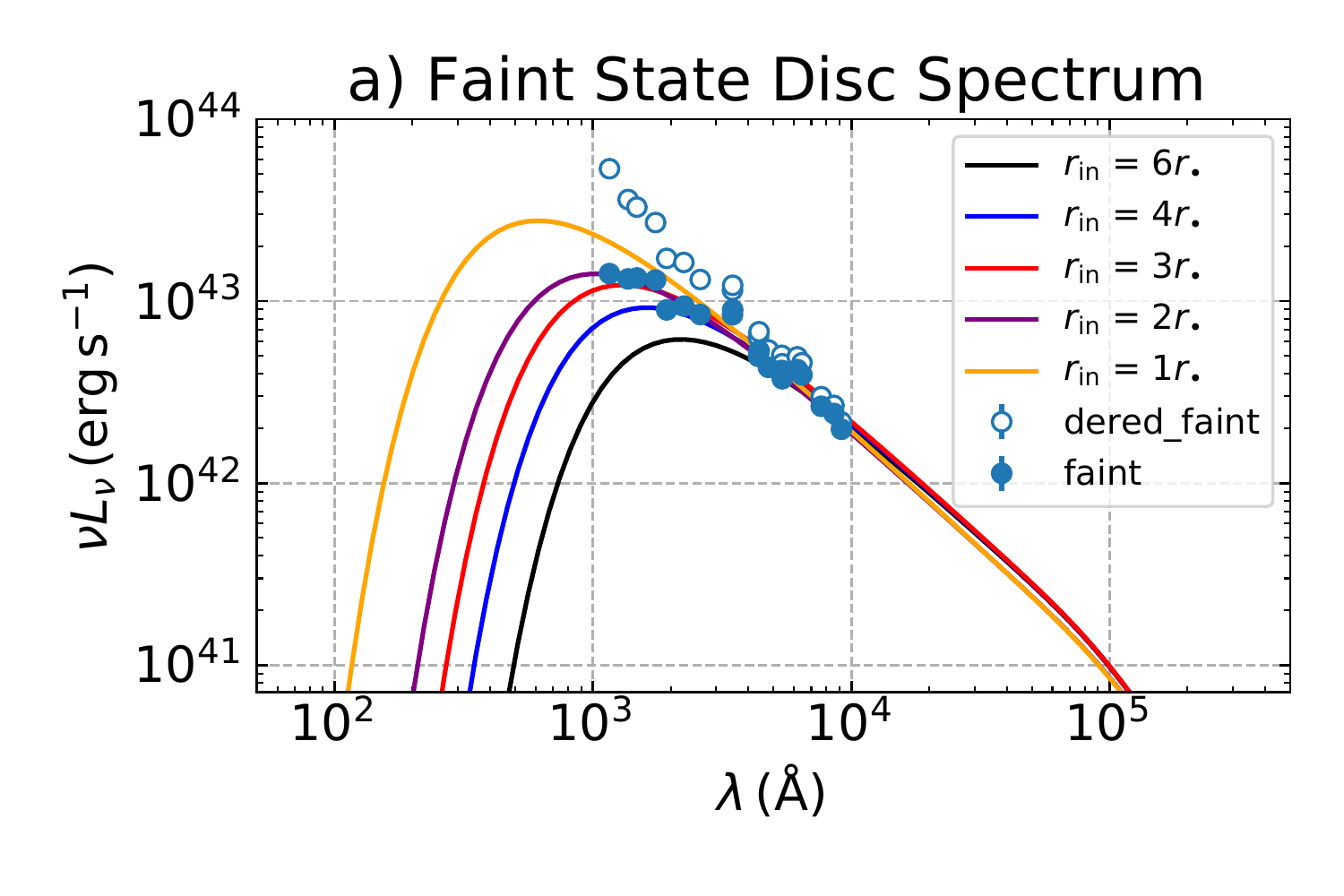}
\\
	\includegraphics[width=0.47\textwidth]{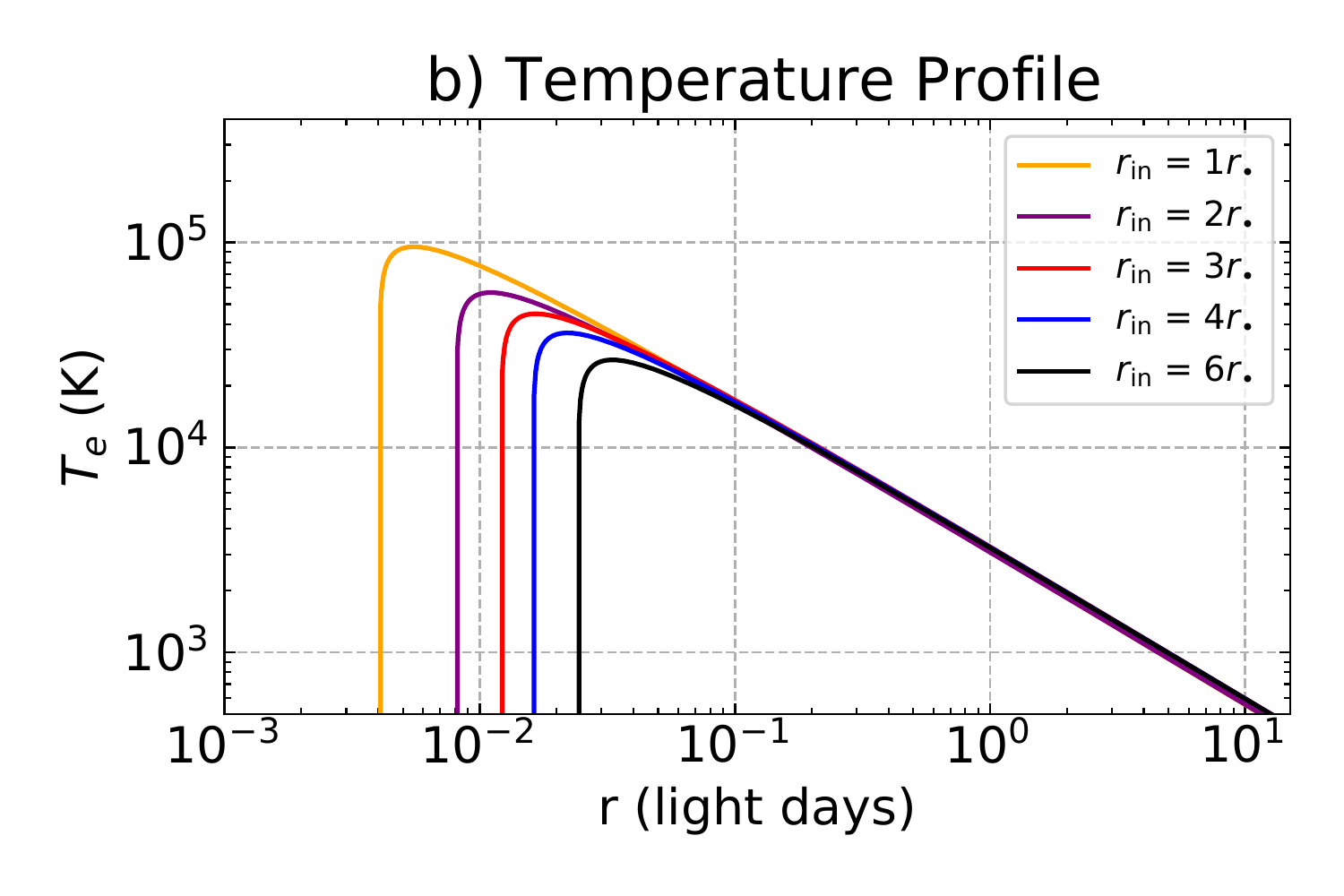}
	\end{tabular}
    \caption{
    \textbf{Panel~a) Disc spectral energy distribution (SED)} $\nu L_\nu(\lambda)$, for 
    models with accretion rate $\dot{M}$ set to match the red end of the observed SED, and with different inner radius $r_{\rm in}/r_\bullet$ = 1, 2, 3, 4, and 6. Filled circles show the observed disc SED for NGC~5548 \citep{st17} at the faintest state of the 2014 STORM campaign. Open circles show the same data after de-reddening using SMC-like dust extinction with $E(B-V)=0.08$. The faint-disc parameters for each model are in Table~\ref{tab:spin}.
    The best fit \change{to the observed faint-disc SED} is for $r_{\rm in}$ = $2\,r_\bullet$. 
    \textbf{Panel~b) Temperature profiles} showing the effective temperatures $T_{\rm e}(r)$ of the disc.
 }
\label{fig:rin}
\end{figure}

\begin{table}
\begin{center}
\caption{ \label{tab:spin}
 Fits to the NGC~5548 faint-disc SED as shown in Fig.~\ref{fig:rin}. 
}
\begin{tabular}{ccccccccc}
\hline
$r_{\rm in}$ 
& $a_\bullet$
& $L_{\rm disc}$ 
& $L_{\rm disc}$ 
& $\dot{M}$ 
& $T_{\rm max}$
& SED fit
\\ 
$r_\bullet$ 
&
& $\dot{M}\,c^2$
& $L_{\rm Edd}$ 
& $M_\odot/{\rm yr}$
& $10^3$~K
& $\chi^2/{\rm dof}$
\\ \hline \hline
6 & 0.00 & 1/12 & 0.0010 & 0.0019 & 
27 & 327/17
\\
4 & 0.57& 1/8 & 0.0015 & 0.0019 & 
37 & 56/17
\\ 
3 & 0.80 & 1/6 & 0.0020 & 0.0019 & 
46 & 23/17
\\ 
2 & 0.93 & 1/4 & 0.0023 & 0.0014 & 
58 & 21/17
\\ 
1 & 1.00 & 1/2 & 0.0045 & 0.0014 & 
96 & 53/17
\\ \hline
\end{tabular}
Notes: $\chi^2$ assumes 0.05~dex ($\sim12$\%) systematic uncertainty. The $\dot{M}$ values are for $M_\bullet=7\times10^7\,M_\odot$ and $i=45^\circ$.
\end{center}
\end{table}

\begin{figure*}
\textbf{ \Large Irradiated flat-disc models}
\centering
	\begin{tabular}{@{}cccccc@{}}
 \multicolumn{1}{l}{}\\
	\includegraphics[width=0.47\textwidth]{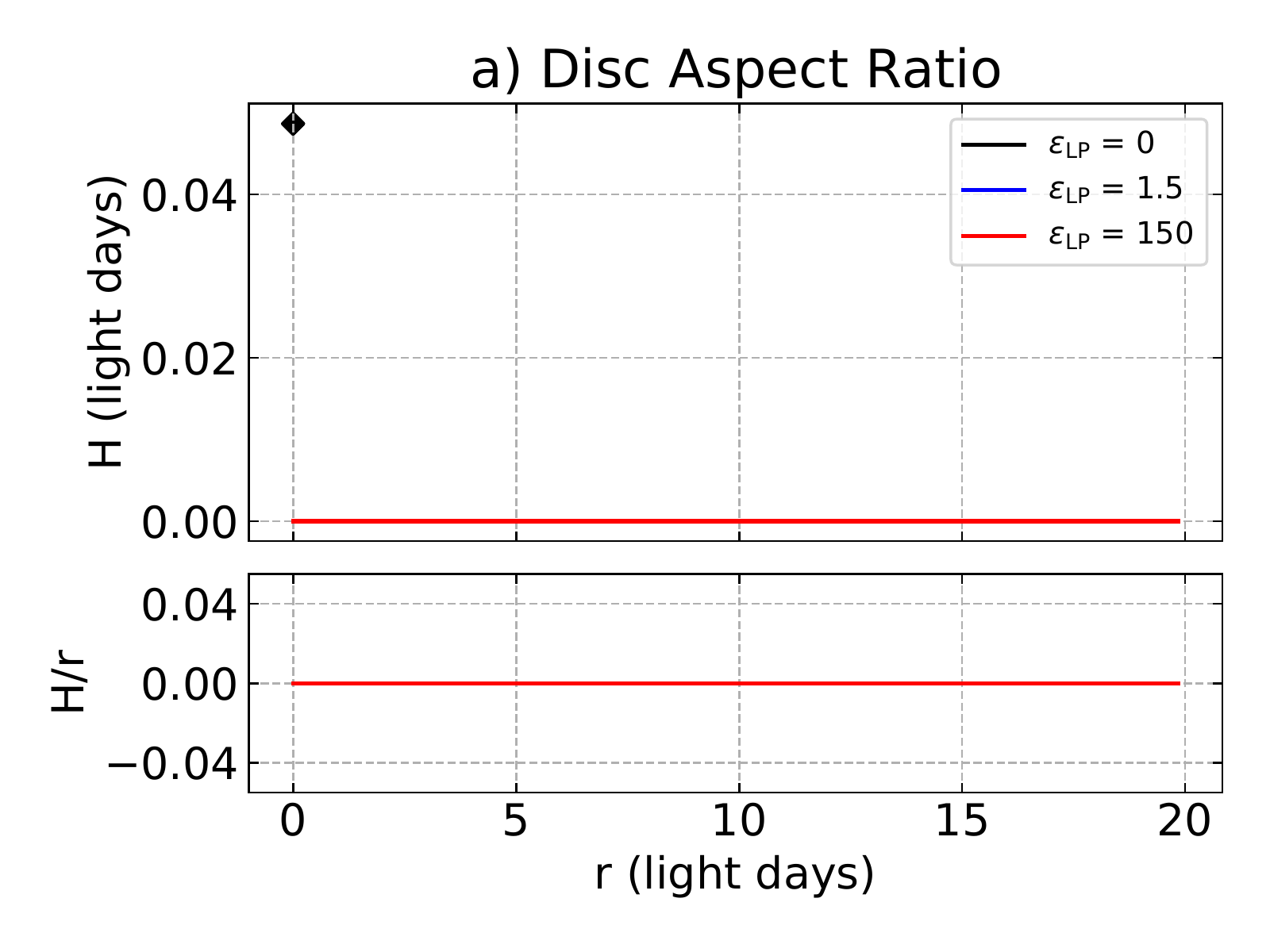}
	\includegraphics[width=0.47\textwidth]{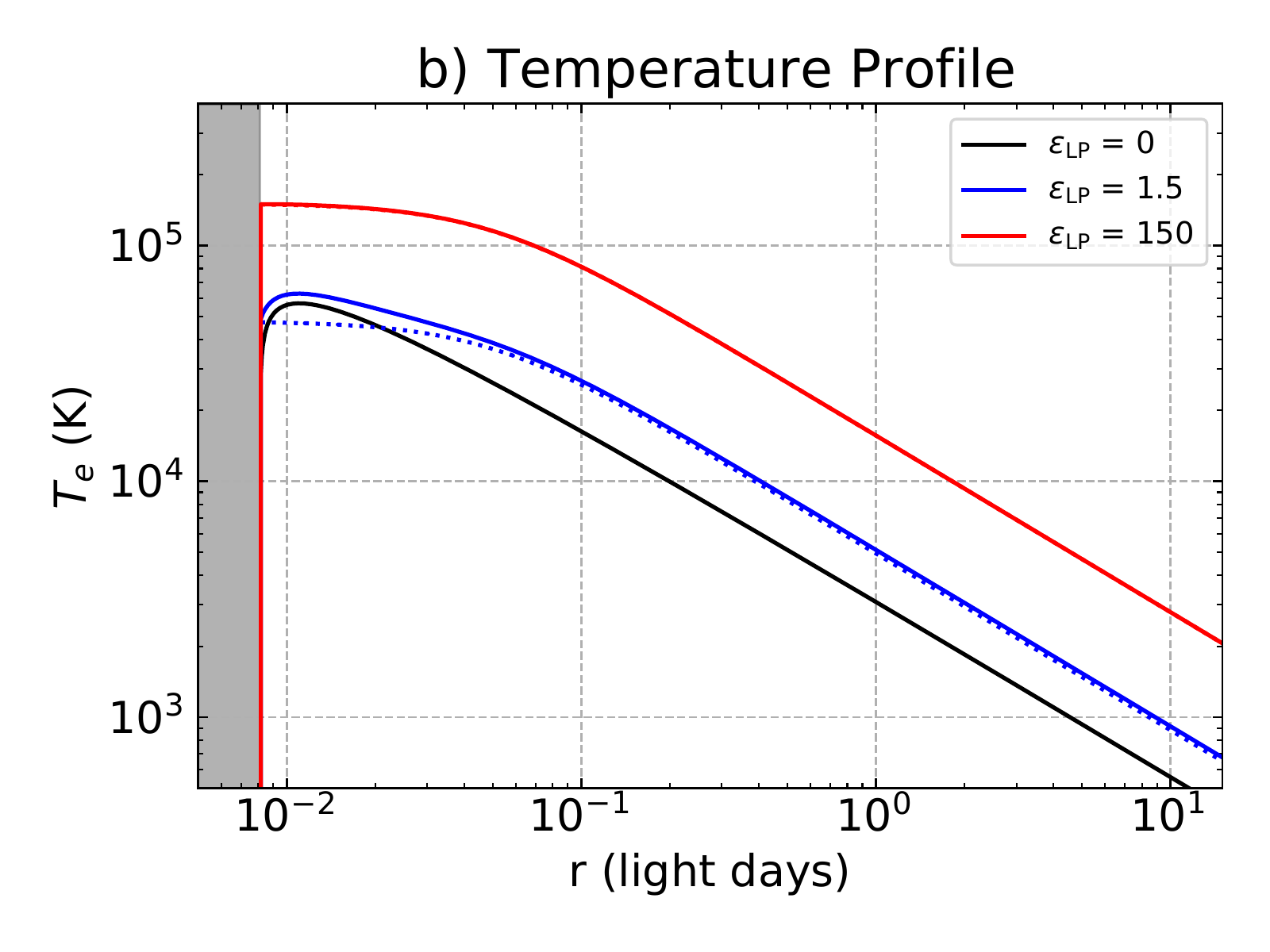}
 \\
  \multicolumn{1}{l}{}\\
	\includegraphics[width=0.47\textwidth]{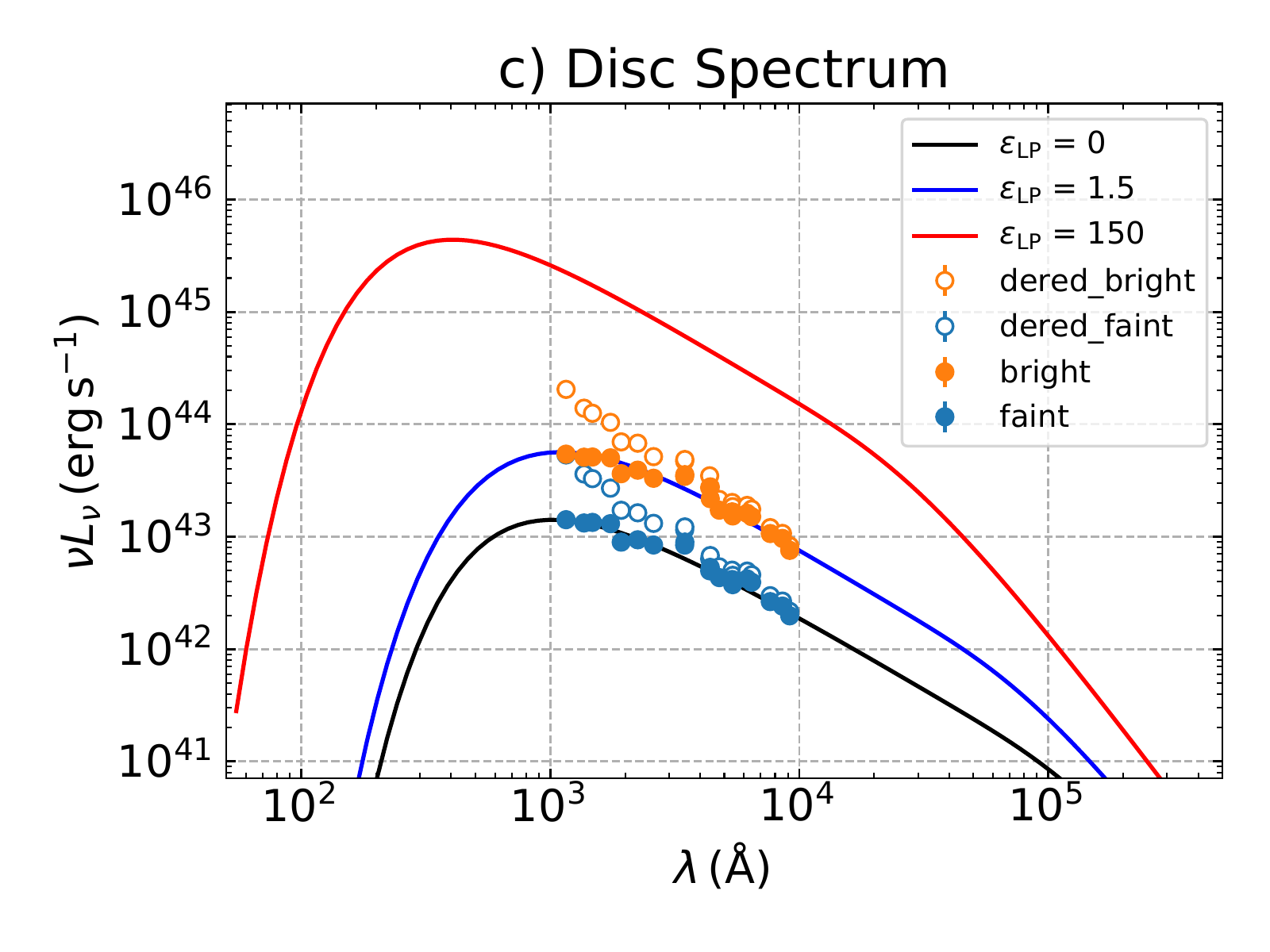}
	\includegraphics[width=0.47\textwidth]{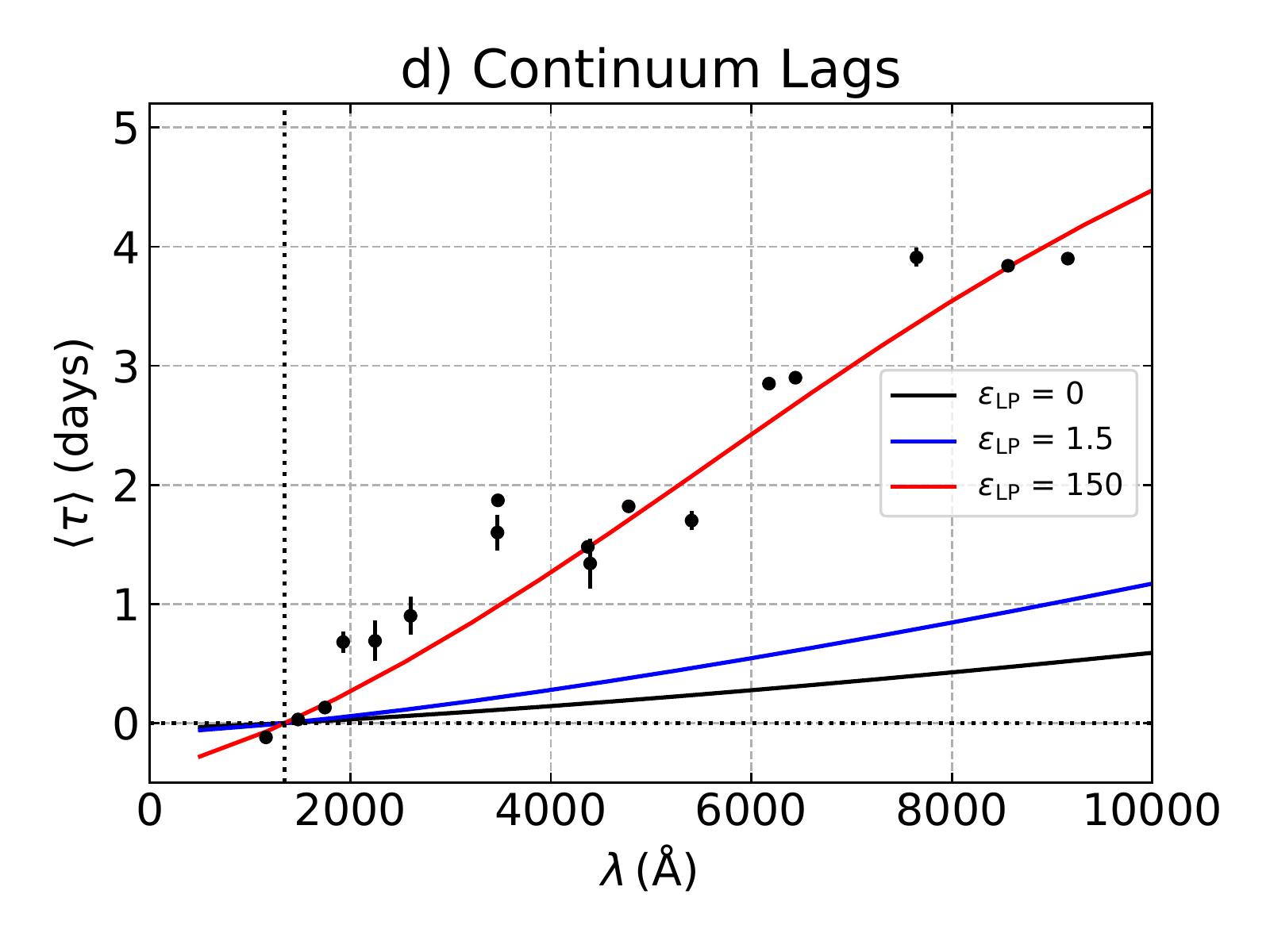}
	\end{tabular}
    \caption{ Irradiated flat-disc models tested by comparison with the continuum time lags and disc SEDs from the STORM campaign on NGC~5548.
    \textbf{Panel~a) Disc thickness profile} for a razor-thin disc, $H(r)=0$, with outer radius $r_{\rm out}=20$~light days and lamp-post height $\Hx=12\,r_{\bullet}\approx 0.05$~light days. 
    \textbf{Panel~b) Temperature profiles} for
    viscous heating alone (black curve, as in Fig.\ref{fig:storm}) and for
     lamp-post irradiation with $\epslp=1.5$ (blue) and 150 (red).
    Solid lines give the total effective temperature, $T_{\rm e}(r)$, balancing both viscous and irradiative heating, and dotted lines give the contribution
    $\Tx$(r) from irradiative heating alone.
 The shaded region indicates radii within the inner-most stable orbit at $r_{\rm ISCO}=2\, r_\bullet$, corresponding
 to a black hole spin parameter $a_\bullet=0.93$.
\textbf{Panel~c) Disc spectral energy distribution (SED)}, $\nu L_\nu(\lambda)$. The faint and bright disc SEDs for NGC~5548 \citep{st17} are shown by the blue and orange circles, respectively, with filled circles for the observed data and open circles for data corrected for SMC-like dust extinction with $E(B-V)=0.08$. 
\textbf{Panel d) Mean lag spectra}, $\taumean(\lambda)$. The error bars show cross-correlation lags \citep{fa16} and dotted lines indicate that the lags are measured relative to the HST 1367\AA\ light curve.
These models adopt a black hole mass $7\times10^7~M_\odot$ and disc inclination $i=45^\circ$ \citep{Horne2021}.
}
    \label{fig:flat_disc}
\end{figure*}

\subsection{Faint-disc SED and accretion rate.}
\label{sec:faintfit}

In the first step, we adjust the accretion rate $\dot{M}$ and the inner radius $r_{\rm in}$ to fit the faint-disc SED.
Our best-fit faint-disc model is the faintest of those shown in Fig.~\ref{fig:storm}, with
system parameters summarised in Table~\ref{tab:faintdisc}.
This faint-disc model has the lamp-post off, 
$\epslp=0$ in Eqn.~(\ref{eq_trprof}), so that the disc is heated entirely by the viscous dissipation and torques associated with the inward accretion flow.
\change{This is a simplifying assumption, since there will be some irradiation in the faint state, and we make it because the faint-state accretion rate and lamp-post irradiation are degenerate parameters.
If we decrease $\dot{M}$, the lamp-post can increase to compensate.
Thus our $\dot{M}$ estimate is an upper limit.}
Note that the shape of the model SED is a good match to the data (black curve in Fig.~\ref{fig:storm}b),
but the predicted
continuum lags are far smaller than
the observed lags (Fig.~\ref{fig:storm}a).
At this stage we ignore the discrepant lags and concentrate on fitting the faint-disc SED.

Fig.~\ref{fig:rin} and Table~\ref{tab:spin} detail several models with different $r_{\rm in}$. All of these have $\dot{M}$ adjusted to fit the IR end of the SED, where
$\nu\,L_\nu\propto(M_\bullet\,\dot{M})^{2/3}\lambda^{-4/3}\cos{i}$. 
In assessing the match between the model and observed disc SED, note that the UV end ($10^3$\AA) is less certain than the IR end ($10^4$\AA).
The UV end is sensitive to possible dust extinction in the host galaxy (open and filled circles in Fig.~\ref{fig:rin}).
Given this uncertainty, we can regard as successful those model discs with SEDs that fall in between the open and
closed circles in Fig.~\ref{fig:rin}.
A good fit on the IR end of the SED
\change{ is insensitive to dust and in our model} requires 
$\dot{M}\approx1.4\times10^{-3}\, M_\odot\,{\rm yr}^{-1}$,
for $M_\bullet=7\times10^7\,M_\odot$ and $i=45^\circ$.

Our estimate for $\dot{M}$
is lower by a factor 5 compared with
previous values
(Table~\ref{tab:faintdisc}) that are
based on fitting the brighter mean disc SED assuming $L_{\rm disk}/L_{\rm Edd}\sim0.1$ and $\epsilon_{\rm disc}\sim0.1$.
We note that the NGC~5548 disc brightens and fades by a factor of 5 over a timescale of a few months, 
much faster than the dynamical time at $r/c\sim5$ light days where the optical continuum arises, and
too fast to be attributed to changes in accretion rate. Accordingly, our $\dot{M}$ is lower \change{than previous estimates} because it corresponds to the faint state, with the lamp-post off.
\change{We emphasise again that our neglect of lamp-post irradiation in the faint state makes our estimate for $\dot{M}$ an upper limit,
and our estimates for lamp-post irradiation are lower limits.}

\subsection{Inner radius and black hole spin.}
\label{sec:spin}

The model SED has a UV downturn corresponding to the maximum temperature $T_{\rm max}$ reached as the inspiraling gas approaches the inner radius $r_{\rm in}$.
We assume that $r_{\rm in}$ occurs at the ISCO radius, which decreases as we increase the black hole spin.
A higher spin has a smaller $r_{\rm ISCO}$, a higher $T_{\rm max}$ and a bluer SED.
To match the UV end of the observed SED, $T_{\rm max}$ needs to be at or above $\sim50\,000$~K.
For a Schwarzshild black hole 
($a_\bullet=0$, $r_{\rm in}=6 \, r_\bullet$), the maximum disc temperature is only $\sim30\,000$~K,
and the model disc SED is too red
(black curve in Fig.~\ref{fig:rin}).
This zero-spin black hole is strongly disfavoured
($\Delta\chi^2=326$).
Instead we need a smaller $r_{\rm in}$ and thus a higher $a_\bullet$.
The model with $r_{\rm in}=3\, r_\bullet$ and
$a_\bullet=0.8$ reaches
$T_{\rm max}=46\,000$~K to give an acceptable fit
($\Delta\chi^2=2)$.
Our best model, with $r_{\rm in}=2\,r_\bullet$ and $T_{\rm max}=58\,000$~K, achieves $\chi^2/{\rm dof}=20.9/17$,
with 2 parameters fitting 19 data.
These $\chi^2$ fit assessments include
a systematic flux uncertainty of 0.05~dex,
about 12~\%,
as determined from the MCMC fit discussed in Section~\ref{sec:mcmc}.
Our best-fit model has a spin parameter $a_\bullet=0.93$, $\epsilon_{\rm disc}=1/4$,
and $L_{\rm disc}/L_{\rm Edd}=0.0023$.

The above spin estimate based on fitting the UV end of the observed disc SED is subject to several caveats.
First, the spin may be higher if we allow for dust.
The maximum spin $a_\bullet=1$ model, with
$r_{\rm ISCO}=r_\bullet$ and correspondingly higher $T_{\rm max}\sim10^5$~K and bluer SED, has a higher $\chi^2$, but is equally acceptable if we allow for possible dust extinction and reddening in the host galaxy.
Second, higher temperatures
may be generated if the inner disc is heated by torques conveyed through magnetic links between the black hole and inner disc \citep{AgolKrolik2000}.
If so, then the spin may be lower than our estimate.
Finally, the lamp-post approximation, neglect of relativity effects, and blackbody assumption may be too simple to realistically describe the UV end of the disc SED.
Nevertheless, the possibility of probing inner-disc physics in this way is an interesting prospect.

\subsection{ Bright-disc SED and lamp-post luminosity. }
\label{sec:brightfit}

Having found a baseline model that fits the faint-disc SED with the lamp-post off, we now turn the lamp on, heating the disc surface, and adjust its intensity until it fits the bright disc SED.
Fig.~\ref{fig:flat_disc} shows,
for the lamp-post model with a razor-thin flat disc with $H(r)=0$ (Panel~a),
the temperature profile (Panel~b) and the corresponding disc SED (Panel~c) and delay spectrum (Panel~d) .
This is similar to Fig.~\ref{fig:storm} except that there the temperature profile is set entirely by viscous heating, with no irradiation, the different models corresponding to different accretion rates.
Here we hold $\dot{M}$ fixed, but increase $\epslp$ in Eqn.~(\ref{eq_trprof})
to turn on the lamp-post. This elevates the disc temperatures, increasing both the disc fluxes and the lags, as shown in Fig.~\ref{fig:flat_disc}. 
The observed SED increases by a factor of 5 without significant change in the spectral shape \citep{st17} The lamp-post heated model
similarly raises the model SED without significantly changing its spectral shape.
The model with $\epslp=1.5$ (blue curve)
achieves a satisfactory fit.

These first 2 steps, fitting the faint and bright disc SED, are successful
with a zero-thickness disc, $H(r)=0$ (Fig.~\ref{fig:flat_disc}c).
In the third step, we aim to fit the continuum lags without upsetting the fit to the bright disc SED. This fails for a zero-thickness disc.
Increasing $\epslp$ helpfully raises both the disc fluxes and the disc lags.
But the model that fits the bright disc SED ($\epslp=1.5$, blue curve) predicts lags that are too small, and the model that fits the lags ($\epslp=150$, red curve) predicts disc fluxes far above those observed.

\doug{The zero-thickness models, though very easy to construct, are inconsistent 
with the requirement of hydrostatic equilibrium in accretion discs heated by either 
viscous dissipation or irradiation.} 
In subsequent sections we address this mismatch by exploring alternative
models that replace the zero-thickness disc
with finite-thickness models that include concave
and convex power-law shapes (Section~\ref{sec:powerlawprofile}) and wave-like ripples (Section~\ref{sec:ripples}) in the thickness profile $H(r)$ of the disc surface. 
{\doug{Since the density in the radiation reprocessing surface is much lower than
that in the midplane, $H(r)$ is likely to be several times larger than the density
scale height in the disc, as inferred for protostellar discs \citep{garaud2007}.}}

\begin{figure*}
\textbf{ \Large Irradiated concave, conic, convex and flat disc models}
\centering
	\begin{tabular}{@{}cccccc@{}}
 \multicolumn{1}{l}{}\\
	\includegraphics[width=0.47\textwidth]{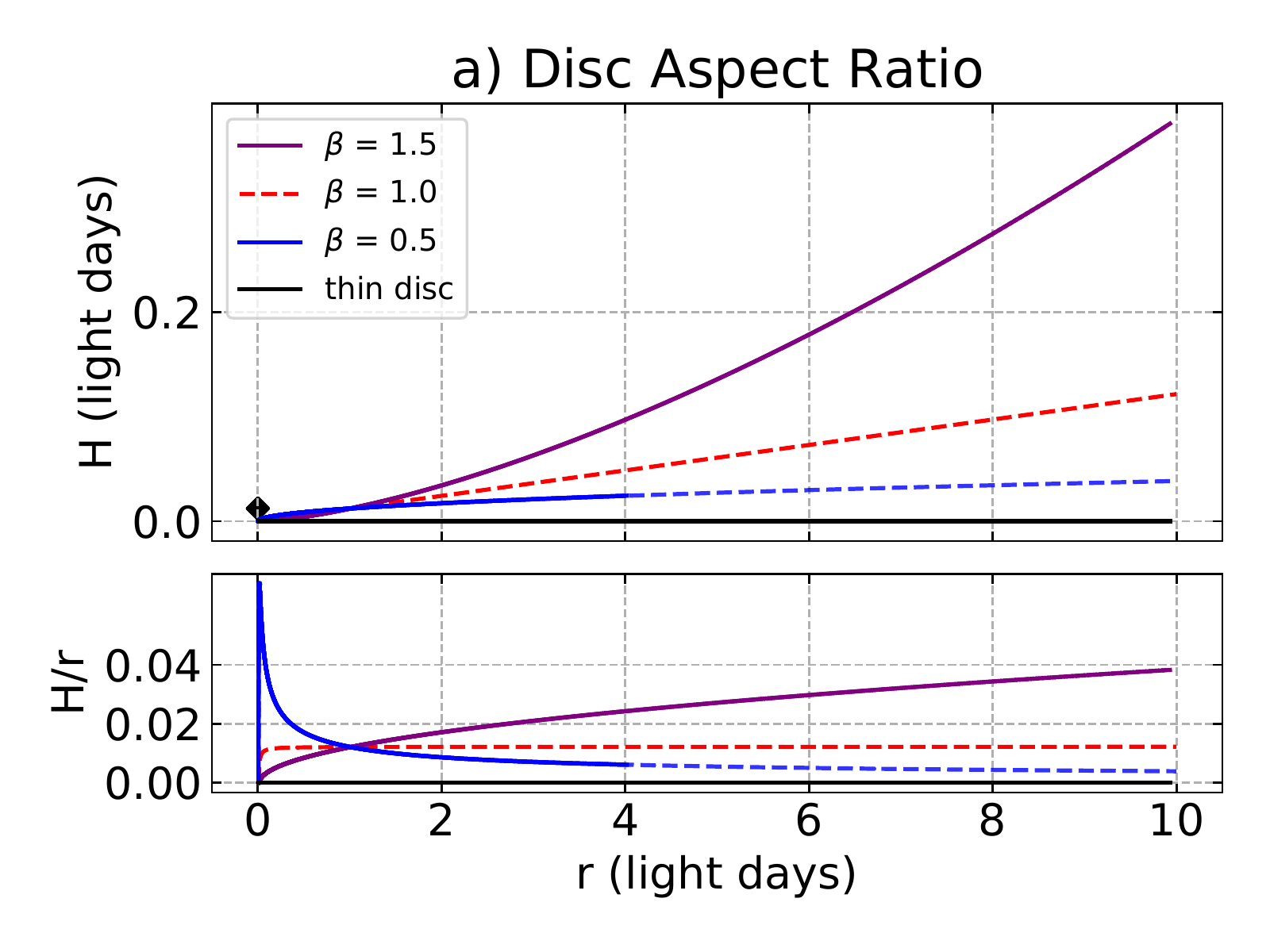}
	\includegraphics[width=0.47\textwidth]{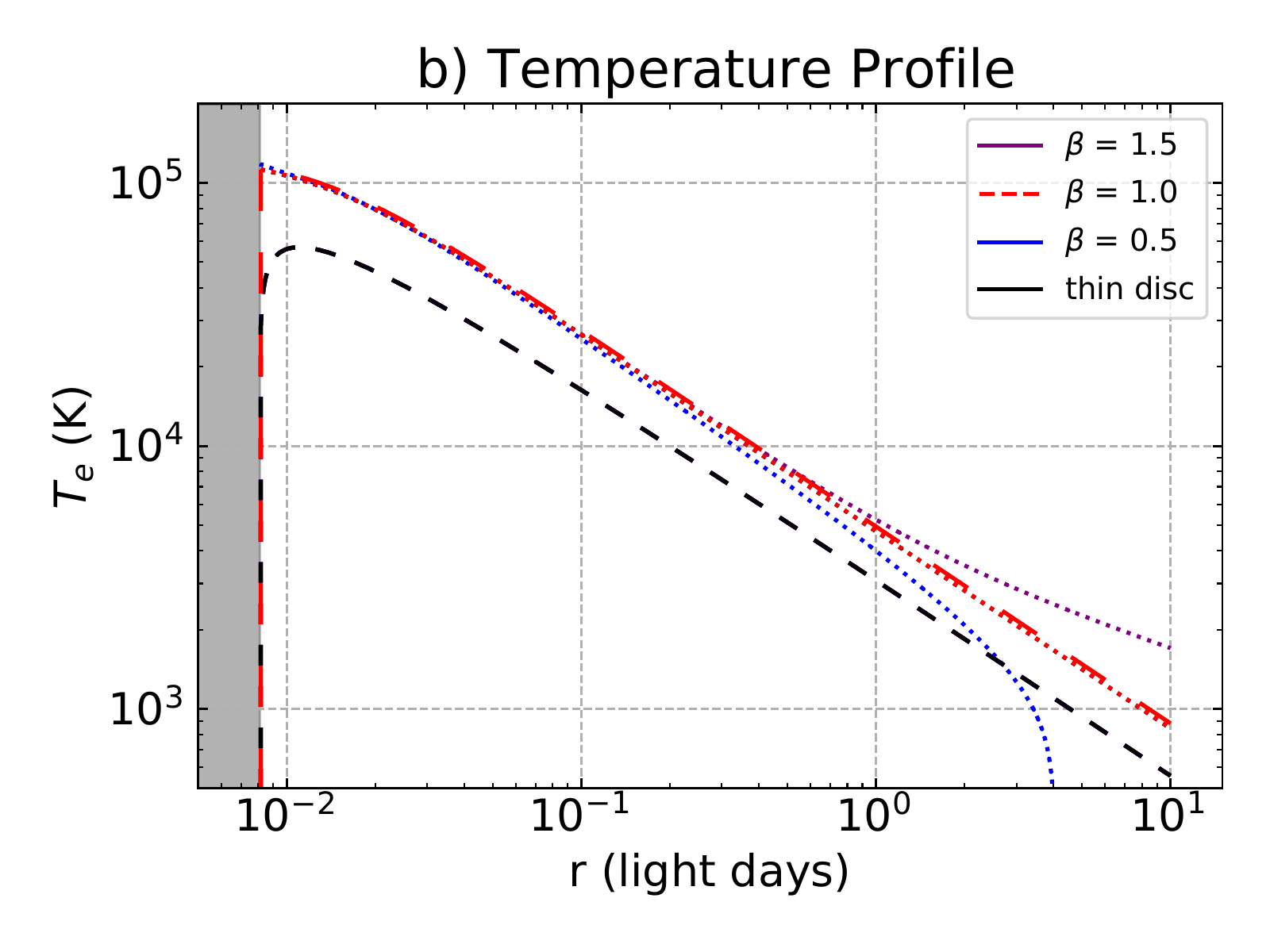}
\\
\multicolumn{1}{l}{}\\
	\includegraphics[width=0.47\textwidth]{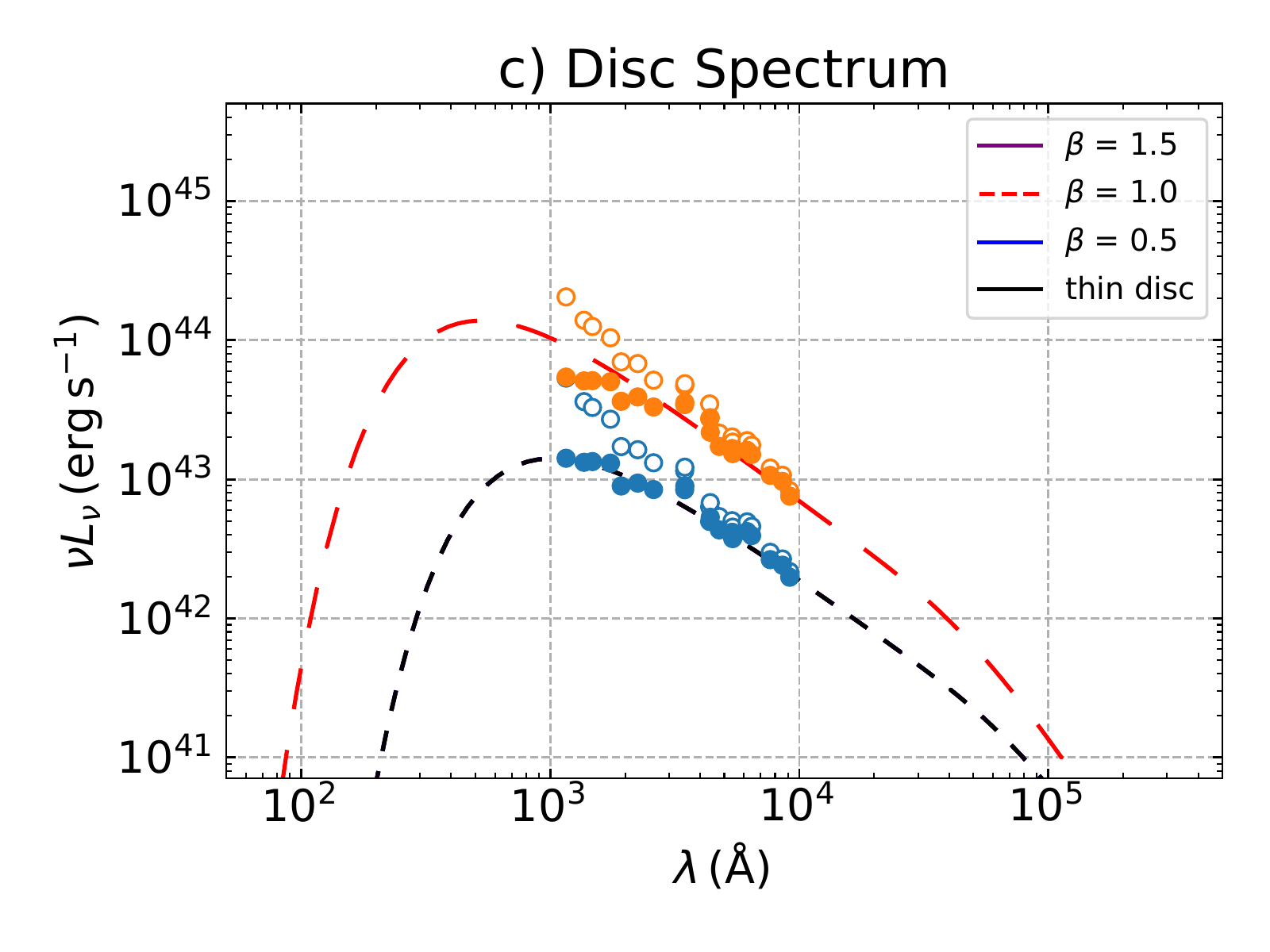}
	\includegraphics[width=0.47\textwidth]{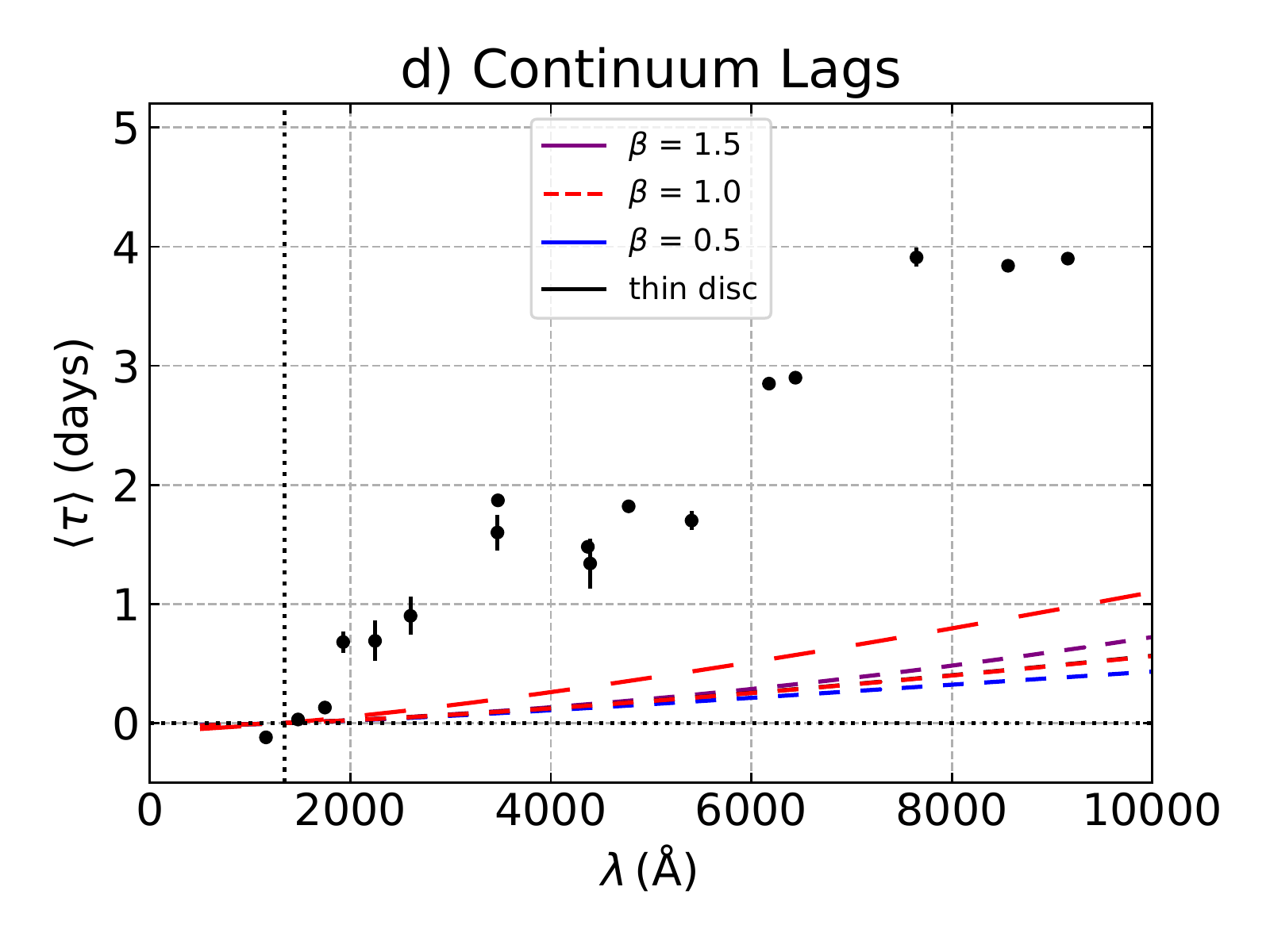}
	\end{tabular}
    \caption{ Similar to Fig.~\ref{fig:flat_disc} but for flat, convex, conic, and concave disc $H(r)$ profiles.
    \textbf{Panel a) Disc thickness profiles} for a razor-thin disc $H(r)=0$ (black),
and for power-law models $H(r)=H_1\,(r/r_1)^\beta$
with thickness $H_1=0.01$ light days at radius $r_1=1$~light days
and power-law indices
$\beta = 0.5$  (blue), 
$\beta = 1$ (red) and 
$\beta = 1.5$  (purple). 
The lamp-post height is $\Hx=3\,R_{\bullet}$.
\textbf{Panel b) Temperature profiles} showing the effective temperatures $T_{\rm e}(r)$ (solid curves) that balance both viscous and irradiative heating, and $\Tx$ (dotted curves)
for irradiative heating alone.
\textbf{Panel~c) Disc spectral energy distributions (SED)}, $\nu L_\nu(\lambda)$. 
With viscous heating alone ($\epsilon_{\rm disc}=
0.25$),
all 4 geometries give nearly identical model
disc spectra (dashed curves) matching the faint-disc data (blue circles).
Turning up the lamp-post ($\epslp=5$), 
all 4 geometries give model SEDs close to
the bright-disc SED
data (orange circles)
from $10^3$ to $10^4$~\AA.
In the infrared ($\lambda>10^4$~\AA),
the flat (black) and conic (red dashed) disc
SEDs are identical, and
the concave (blue) and convex (purple) disc
SEDs are bluer and redder, respectively.
\textbf{Panel~d) Mean lag spectra}, $\taumean(\lambda)$. Solid lines trace mean lag spectra for the bright-disc models ($\epslp =5$), and dashed lines for the same models with lamp-post turned off ($\epslp =0$). 
All models fall short of the observed lags, but the concave geometry moves in the right direction.
}
    \label{fig:powerlaw}
\end{figure*}

\section{Discs with power-law $H(r)$ profiles}
\label{sec:powerlawprofile}

To address the mismatch between observed
lags and those from our disc models,
while maintaining their good fit to the 
faint-disc and bright-disc SEDs,
 we generalise the 
zero-thickness disc model, $H(r)=0$, to a finite thickness power-law model:
\begin{equation}
\label{eq_hpowerlaw}
	H(r) =  H_1 \left(\frac{r}{r_1}\right)^\beta
	\ .
\end{equation}
\noindent Here $H_1$ is the disc height at fiducial radius $r_1$ and $\beta$ is the power-law slope.
This parameterisation allows for concave ($\beta>1$) and convex ($\beta<1$) surface profiles that modify the way in which the 
central lamp-post irradiates the disc surface at each radius, as well as altering the light travel time delays.

A power-law disc thickness profile can be constructed theoretically from a standard disc model with an appropriate opacity law
in the energy equation. In the relevant regions of AGN discs (with $r \sim$ a few light days) where 
radiation pressure and electron-scattering  opacity  dominate, 
$\beta = 9/8$ \citep{ss73, AP}.
(Interior to this region, where gas pressure dominates, $\beta \simeq 0$.) 

Fig.~\ref{fig:powerlaw} compares predictions for this model, relative to the thin-disc predictions, for power-law indices $\beta=0.5$, $1.0$, and $1.5$, corresponding to convex, conic, and concave disc surfaces, respectively. 
The lamp-post luminosity is calculated using $\epslp=5$.
The format of Fig.~\ref{fig:powerlaw} is the same
as that in Fig.~\ref{fig:flat_disc}.
Panel~a shows the model disc thickness $H(r)$
and
Panel~b the corresponding effective temperature profile $T_{\rm e}(r)$.
Panel~c shows model disc SEDs,
$\nu\,L_\nu(\lambda)$,
along with the faint-disc and bright-disc SED data obtained during the STORM campaign on NGC~5548.
Panel~\ref{fig:powerlaw}d shows the model delay spectra $\taumean(\lambda)$ along with the measured cross-correlation lags. 
Note here that the concave disc surface ($\beta=1.5$) increases the lags, 
and the convex disc surface ($\beta=0.5$) decreases the lags,
as discussed in more detail below.

Considering first a face-on viewing angle, $i=0$, the concave model ($\beta>1$) reduces lags at longer wavelengths by elevating the disc (thus reducing the light travel time lag) and also by tilting the disc surface toward the lamp-post (thus increasing the irradiation).
Conversely, a convex $H(r)$ profile ($\beta<1$) reduces irradiation in the outer regions,
steepening the $T_{\rm e}(r)$ profile.
Both effects alter the predicted delay spectrum $\tau(\lambda)$ in ways that may
reduce tension between the observations and the model predictions.


 The temperature profile for such a model is given by Eqn.~(\ref{eq_trprof}) 
 with a covering factor of the form
\begin{equation}
\label{eq_covfracpowerlaw}
	f(r) = \frac{ \Hx }{ r }
	+ \left( \beta - 1 \right) \frac{H_1}{r} \left( \frac{r}{r_1}\right)^\beta 
= \frac{ \Hx }{ r }\, \left[ 1 - \left( \frac{ r }{ r_\beta } \right)^\beta \right] \ .
\end{equation}
where $r_\beta = r_1 ({\Hx }/{  H_1\, (1 - \beta)})^{1/\beta}$.

For both flat ($\beta=0)$ 
and  conic ($\beta=1$) disc $H(r)$ profiles,
which are relevant for the gas and radiation pressure dominated, non-self-gravitating, inner region of AGN discs,
the $f(r)$
expression in Eqn.~(\ref{eq_covfracpowerlaw})
reduces to the thin-disc case, Eqn.~(\ref{eq_covflat}). We
see this property in Fig.~\ref{fig:powerlaw} where the flat disc with $\beta = 0$
(black line) and conic disc with $\beta = 1$ (red dashed line) exhibit the same temperature profile,
$T_{\rm e}\propto r^{-3/4}$,
the same SED, $\nu\,L_\nu\propto\lambda^{-4/3}$,
and the same continuum lags $\tau\propto\lambda^{4/3}$.

For convex disc $H(r)$ profiles ($\beta < 1$), the accretion disc at radius $r$ must satisfy the condition 
\begin{equation}
\label{eq_powerlaw_expose}
	f(r) > 0 \ \ \ \ {\rm or} \ \ \ \ r \leq r_\beta
\end{equation}
in order to be exposed to the lamp-post irradiation. 
This condition is satisfied if 
\begin{equation}
 \frac{ \Hx }{H(r)} = \frac{ \Hx }{ H_1} \,\left( \frac{r }{ r_1} 
\right)^{- \beta} > 1-\beta.
\label{eq:betacon}
\end{equation}
In this region, 
the magnitude of $f$ decreases with $r$, Eqn.~(\ref{eq_covfracpowerlaw}) substitutes into Eqn.~(\ref{eq_trprof}) to give 
\begin{equation}
	T_{\rm e}^4 = \left( \frac{G \,M_\bullet \,{\dot M} }{ 2 \, \pi \, \sigma \, r^3} \right)\,
	\left(
	\frac{ 3 }{ 4 }
	+ \frac{ \epslp \, \Hx }{ r_\bullet } 
	\left[ 1 - 
	\left( \frac{ r }{ r_\beta } \right)^\beta 
	\right] 
	\right)
	\ ,
\label{eq:fbeta}
\end{equation}
so that $T_{\rm e}$ decreases with $r$ faster than $r^{-3/4}$
and $\nu L_\nu$ decreases with $\lambda$ faster than $\lambda^{-4/3}$.
Outer regions of the disc (at $r> r_\beta$, where $f<0$) are ``in the shadows'' of the inner regions and thus are not heated by irradiation. 
\noindent 

For convex disc $H(r)$ profiles ($\beta<1$),
the outer regions can be
in the shadow of the inner regions, i.e. not exposed to the lamp-post, so that
$T_{\rm e} \left( r \right) \simeq T_{\rm v} = ({3\,G\,M_\bullet \,\dot{M}}/{8 \,\pi\, \sigma \, r^3 })^{1/4}$.
The blue curves in Fig.~\ref{fig:powerlaw}, for $\beta=0.5$, show an example in which 
the outer regions ($r>r_\beta\approx 4$~light days) are 
in the shadow of the inner regions.
This shadowing of the cooler outer-disc regions makes the SED bluer (Fig.~\ref{fig:powerlaw}c) and decreases the lags (Fig.~\ref{fig:powerlaw}d).
 
For concave disc  $H(r)$ profiles ($\beta > 1$), Eqn.~(\ref{eq_powerlaw_expose}) is always satisfied,
i.e. all parts of the disc are exposed to the irradiation (red and purple lines in Fig.~\ref{fig:powerlaw}).  
The irradiative heating term, $\epslp\,(r/r_\bullet)\,f(r)$ in Eqn.~(\ref{eq_trprof}), is an increasing function of $r$ 
so that $T_{\rm e}$ decreases with $r$ less steeply than the $r^{-3/4}$ (Fig.~\ref{fig:powerlaw}b),
the SED becomes redder (Fig.~\ref{fig:powerlaw}c)
and the lags increase (Fig.~\ref{fig:powerlaw}d).

\begin{figure*}
\textbf{ \Large Irradiating a flat disc with a steep rim}
\centering
	\begin{tabular}{@{}cccccc@{}}
 \multicolumn{1}{l}{}\\
	\includegraphics[width=0.47\textwidth]{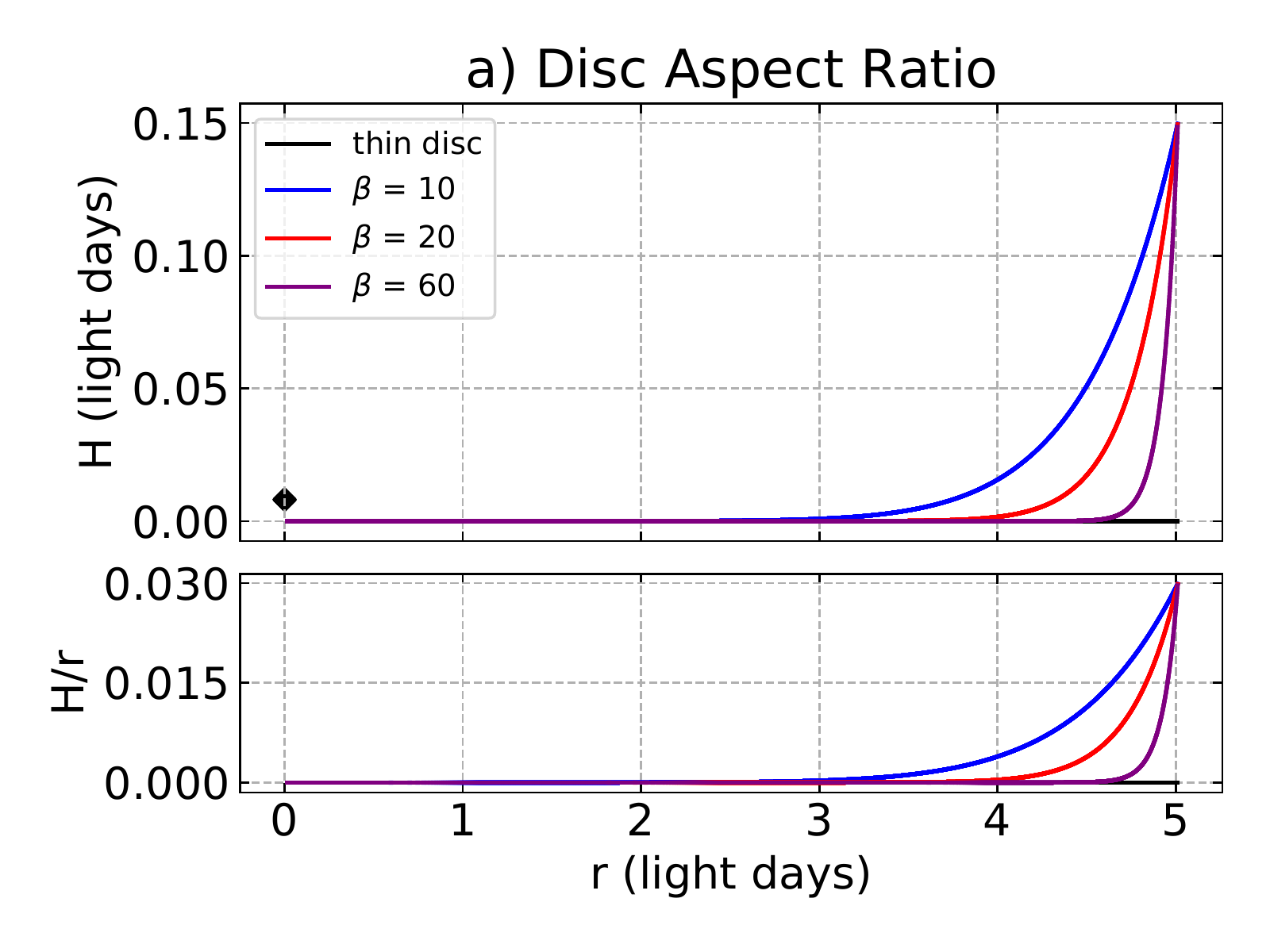}
	\includegraphics[width=0.47\textwidth]{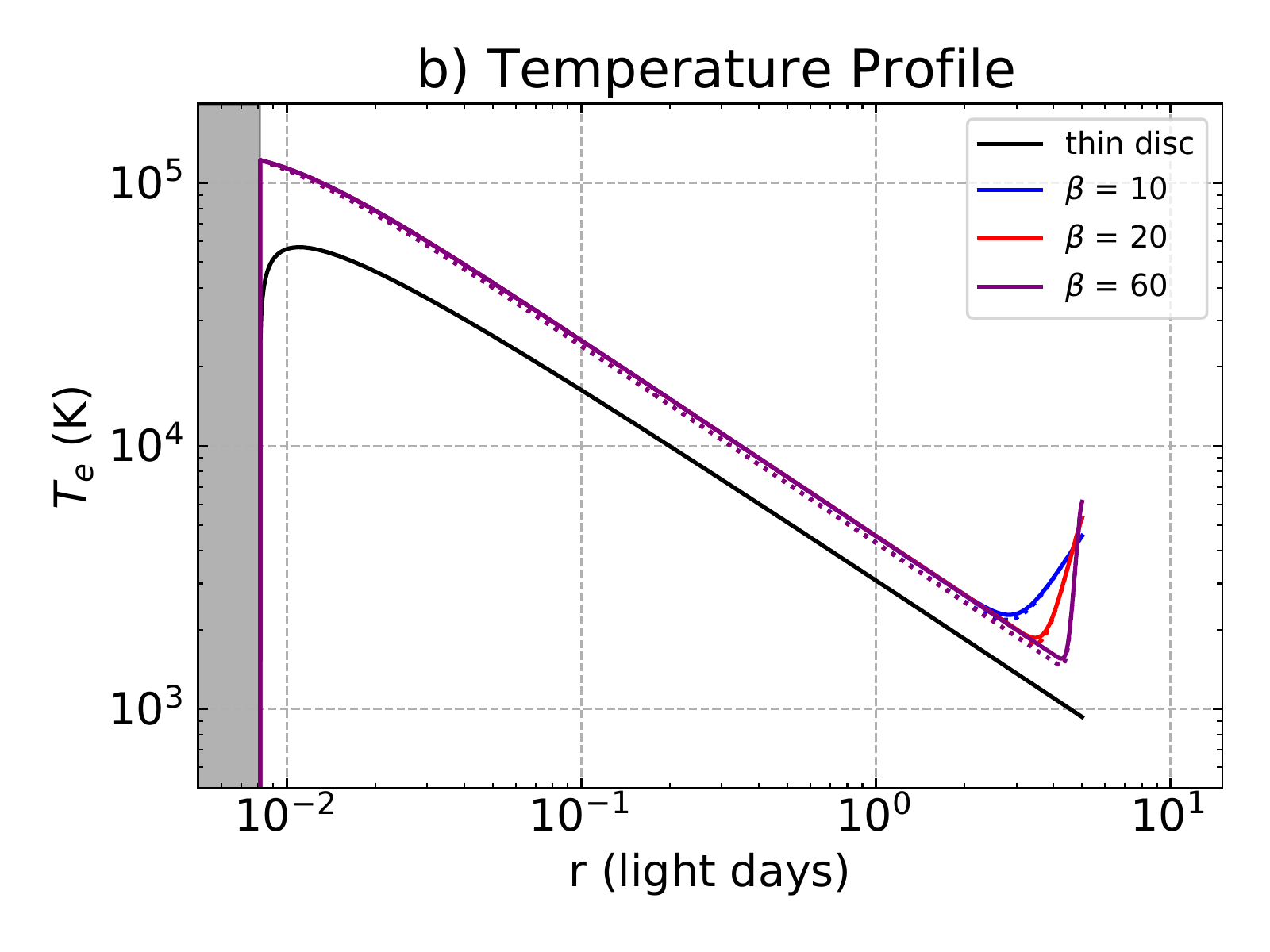}
 \\
  \multicolumn{1}{l}{}\\
	\includegraphics[width=0.47\textwidth]{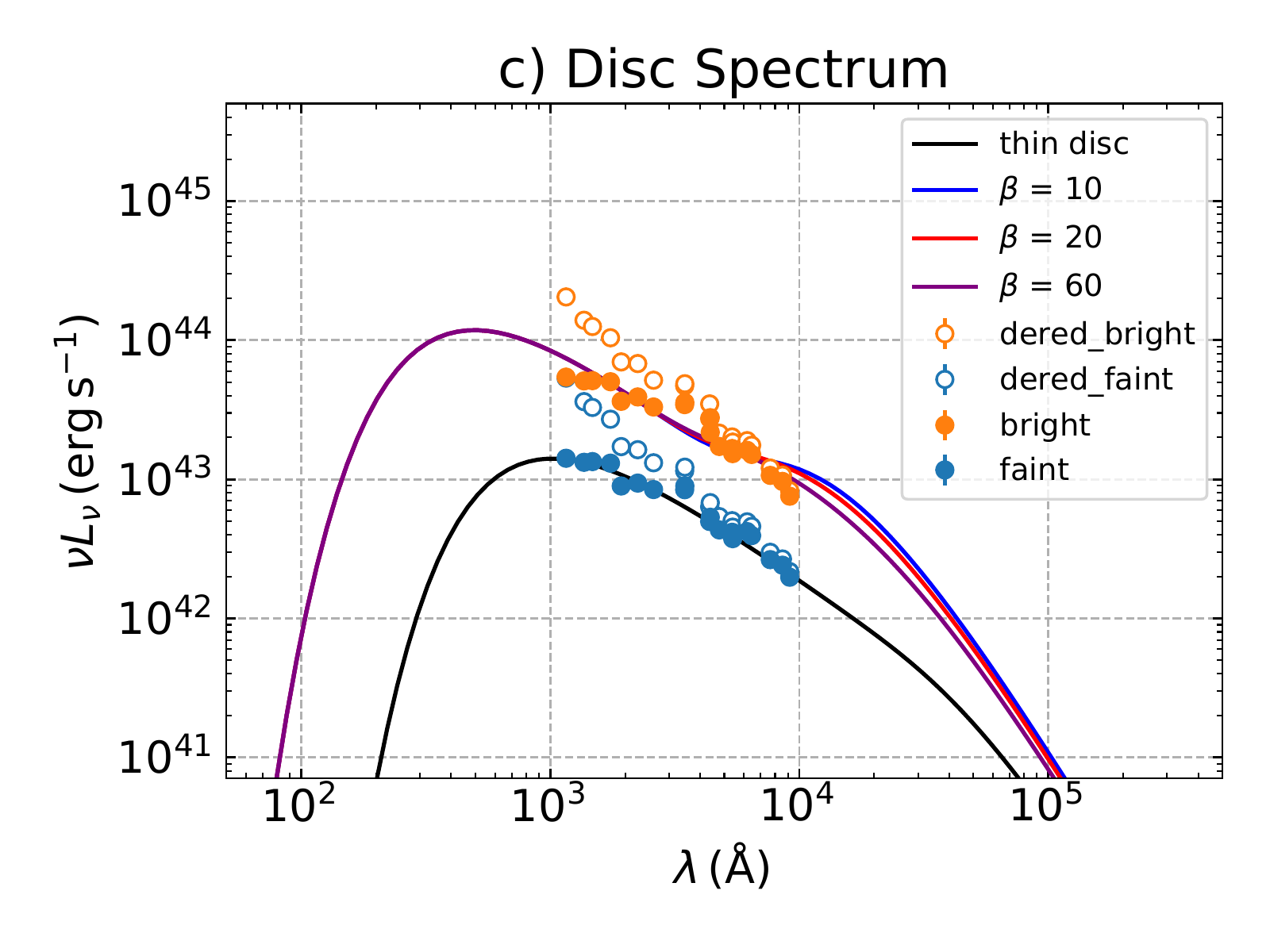}
	\includegraphics[width=0.47\textwidth]{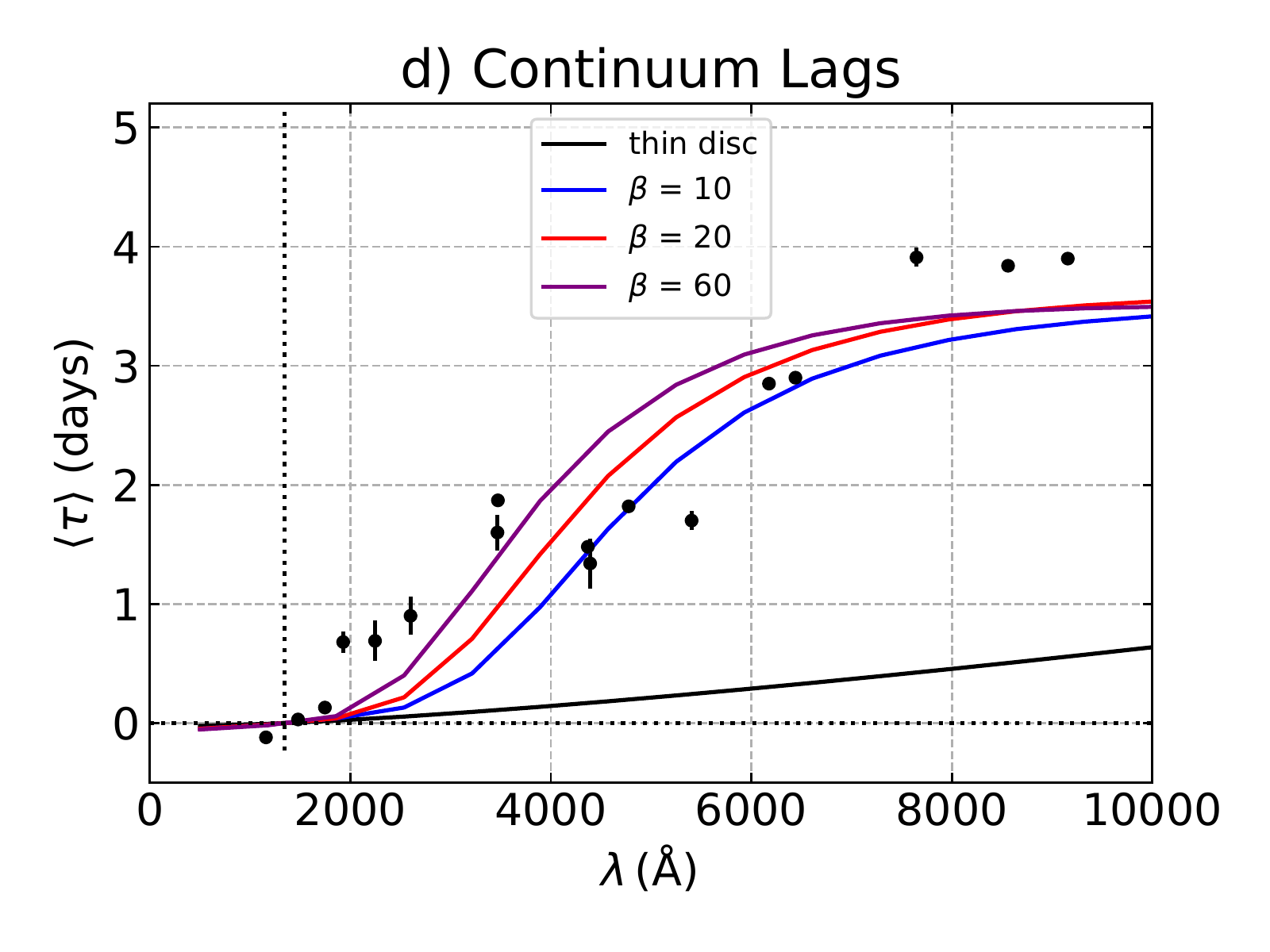}
	\end{tabular}
    \caption{Similar to Figs.~\ref{fig:flat_disc} and \ref{fig:powerlaw} but for nearly-flat discs with a relatively steep rim that rises to $H_1= 0.15$~light days ($H/r=0.03$) at the outer radius $r_1=5$~light days
    (Panel~a).
    The disc height is 
    $H(r)=H_1\,\left(r/r_1\right)^\beta$ with a razor-thin disc $H(r)=0$ (black),
 and increasingly-steep rims produced with
$\beta = 10$  (blue), 
$\beta = 20$ (red) and 
$\beta= 60$  (purple). 
The effective temperature profile (Panel~b) falls as $T_{\rm e}\propto r^{-3/4}$ on these nearly-flat
discs, reaching $10^3$~K at $r=r_1$
with the lamp-post off.
 Turning the lamp-post efficiency up to $\epslp=5$, the temperature rises by a factor $\sim1.5$ on the nearly-flat parts of these discs,
 and rises rapidly to $\sim6000$~K due to irradiation on the steep inner face of the rim.
 The model disc SEDs (Panel~c) match the
 faint-disc data (blue circles) with the lamp-post off,
 and the bright-disc data (orange circles) with the lamp-post on.
 For these models with fixed aspect ratio ($H/r=0.03$) at $r=5$ light days, steeper rims have a smaller rise in flux at $10^4$~\AA, and models with $\beta > 60$ have spectral slopes that match the data.
With the inclined viewing angle $i=45^\circ$,
the steep rims dramatically increase the model
lags (Panel~d) bringing them close to the data.
These steep-rim disc models achieve a satisfying fit to both the continuum lag and disc SED data.
}
    \label{fig:steep}
\end{figure*}

 To summarise the above considerations, illustrated in Fig.~\ref{fig:powerlaw}, the shape of the disc thickness profile $H(r)$ can alter 
 the irradiated disc's $T_{\rm e}(r)$ profile
in ways that are helpful in moving the model predictions closer to the observed disc lags. 
In particular, a concave disc surface helpfully increases the lags, but also increases fluxes on the red end of the SED.
The key issue then is to retain the larger lags without producing an excessive increase in the red flux.

\subsection{Fits with a concave bowl-like $H(r)$ profile}
\label{sec:bowl}

To reconcile the conflicting constraints from
the lags and the SED, a promising option is
a concave bowl-shaped disc surface.
This exposes annuli at larger radii to a larger solid angle of lamp-post irradiation, increasing $T_{\rm e}(r)$ in the outer disc.
For the power-law profiles considered in Eqn.~(\ref{eq_hpowerlaw}),  
$\beta>1$ produces a concave disc surface on which
the power-law slope of $T_{\rm e}(r)$ is $-3/4$ in
the inner disc where $H(r)\ll \Hx$, and less steep
($> -3/4$) in the outer disc where $H(r)\gg \Hx$.
We need to boost temperatures to $\sim6000$~K
at $r/c\approx5$ days to increase the
lag at 9000~\AA\ toward the maximum lag
$\tau_{\rm max}\approx(r/c)\,(1+\sin{i})$.

A second effect that helps to increase lags is the result of our viewing angle, $i=45^\circ$,
looking into the concave bowl shape of the
irradiated disc. As the bowl is tipped away from a face-on view, the near side becomes less visible than the far side
as a result of the foreshortening of disc surface elements.
Thus relative to a flat disc, for which the mean lag is independent of inclination, we see
a larger response on the
far side of the tilted bowl relative to the near side, and this increases the lags.

A drawback of the bowl-shaped disc surface
is that while it increases the lags it
also increases the flux on the red end of the SED.
This can already be seen 
in Fig.~\ref{fig:powerlaw}c for the $\beta=1.5$ model.
This unwanted red flux excess can be reduced by decreasing the outer disc radius $r_{\rm out}$,
but that also reduces the lags.
A compromise is needed between fitting the lags and fitting the SED.

We attempted to realise this compromise by
adjusting the depth and shape of the bowl, the height of the lamp-post, and the outer radius, with some success.
But a fully optimised model was hard to discover while moving around in this 4-dimensional parameter space. The best result we achieved by hand was for a model with $\beta=1.4$, but with a relatively thick disc, $H/r\sim0.2$ at $r=r_{\rm out}\sim10$ light days.
This ``slim disc'' model comes close to fitting the data, but is less natural than a thinner disc model in which vertical hydrostatic equilibrium gives a smaller $H/r$, of order the ratio of sound speed to Kepler velocity.

\subsection{MCMC fits yield a thin disc with steep rim}
\label{sec:mcmc}

After some investigations adjusting parameters by hand, we automated the optimisation of 4 model parameters, the lamp-post luminosity
$\Lx$, the disc thickness and shape
parameters $H_1$, $\beta$, and the outer disc radius $r_{\rm out}$, with fixed $\Hx$.
We implemented Markov-Chain Monte Carlo (MCMC) sampling to search the parameter space of the model and improve the initial fit.
Replacing $H_1$ at $r=r_1/c=1$ day
by $H_{\rm out}$ at $r=r_{\rm out}$ 
provided a more nearly orthogonal set of parameters.
To keep all 4 parameters positive, we adopted priors uniform in their log.
Assuming Gaussian errors on the flux and lag measurements, we adopt the usual
the badness-of-fit metric,
\begin{equation}
 -2\ln{L} = 
 \chi^2 
 + \sum_{i}\ln{\left(V_i\right)}
 \ .
 \end{equation}
 Here $\chi^2=\sum_i(D_i-\mu_i)^2/V_i$,
 sums the squares of fit residuals between the data $D_i$ and corresponding model $\mu_i$, normalised to the
 variances $V_i = \sigma_i^2 + \sigma_0^2$.
 The sum extends over both the lag and flux data,
 18 lags, and 19 fluxes in the bright and faint states, a total of 56 data.
 We include a noise-model parameter $\sigma_0$, added in quadrature with the nominal flux errors $\sigma_i$. This extra variance parameter enables the model to relax its attention to the SED data and thereby to improve the fit to the lag data. Thus we adjust 5 parameters to fit 56 data.

 The result of our MCMC fitting, from several starting points in the parameter space, was to drive $\beta$ up to large values,  $\beta>100$. 
 This automated search produces 
 a disc geometry that is very flat out to $r/c\sim4$ days, with a steep rise at $r_{\rm out}/c\approx 4$ days. 
 The model SED is dominated by the flat inner disc, fitting the faint-disc and bright-disc SEDs
 well. The flux noise parameter is optimised at $\sigma_0\approx 0.05$~dex, about 12\%.
 
 Fig.~\ref{fig:steep} shows 3 models with increasing $\beta$ producing a steep rim on the outer edge of the disc, here set to $r_{\rm out}=5$~light days (Panel~a).
 The rim height is $H_{\rm out}=0.15$~light days, corresponding to
 $H/r=0.03$.
 With a modest $\epslp=5$, the declining $T_{\rm e}\propto r^{-3/4}$ on the flat inner disc 
 is elevated by a factor $\sim1.5$, and rises
 rapidly to $T_{\rm e}\sim6000$~K at the irradiated rim (Panel~b).
 The steep rim contributes little to the SED (Panel~c),
 and the unwanted red flux actually decreases as the rim becomes steeper.
 However, the steep rim greatly increases the lags (Panel~d),
 which now rise with wavelength to 3.5 days,
 coming much closer to the observed lags $\sim4$~days.

To understand this behaviour, note that the steep rim at $r=r_{\rm out}$ produces a U-shaped delay distribution, with a sharp peak at $\tau_{\rm max} \approx (1+\sin{i})\,r_{\rm out}/c$
 from the far side, and a weaker peak
 at $\tau_{\rm min} \approx (1-\sin{i})\,r_{\rm out}/c$
 from the near side.
 The delay distribution $\Psi(\tau)$ is particularly simple for an infinitely steep vertical rim ($\beta=\infty$),
 $\Psi=0$ for $\tau<r_{\rm out}/c$, since the near half of the rim faces away from the observer, then rising linearly, $\Psi\propto(\tau-r_{\rm out}/c)$, to a maximum at $\tau=\tau_{\rm max}$.
 With this distribution
 the mean lag from a vertical rim is 
 \be
 \left<\tau\right>_{\rm rim}
    = \frac{r_{\rm out} } { c }
    \, \left( 1 + \frac{2}{3}\, \sin{i} \right)
    \ .
 \ee
 Inside the steep rim, the flat disc with $T_{\rm e}\propto r^{-3/4}$ produces a relatively prompt 
 response with a lag increasing as $\tau_{\rm disc}=\tau_0\,\left(\lambda/\lambda_0\right)^{4/3}$.
 The combined disc + rim response has a lag in between these extremes, moving from one to the other as their relative contributions change with wavelength.
 The flat disc response dominates at UV and blue optical wavelengths and the rim response becomes increasingly prominent at redder optical wavelengths.
 
 The result is a plausible fit to the bright and faint disc SEDs and simultaneously to the lags, as
 shown in Fig.~\ref{fig:steep}.
 The shape of the model lag spectrum is not perfect, however.
 While the model lags are small in the UV, rising rapidly in the optical and becoming flat in the IR, the observed lags increase more nearly linearly with wavelength.
 Nevertheless, this flat disc + steep rim model
is the first model we found that achieves a plausible fit to both the lags and the SEDs. 

 A steep rim may arise from a tearing
 structure in a warped disc \citep{nixon2012} 
 or the launch pad of a failed disc wind
 \citep{czerny2011, baskin2018, naddaf2021}. 
The failed disc wind model 
\changes{
was proposed as a mechanism for 
the origin of the broad line region (BLR).
} 
It is particularly compelling here, since it predicts an increase 
in the disc thickness when the surface temperature falls below the dust sublimation temperature, about 1000~K 
for carbonaceous dust or 1500~K for silicate dust. Notice in Fig.~\ref{fig:steep}b that the disc temperature 
just inside the steep rim is indeed around 1500~K, compatible with silicate dust. The dust opacity intercepts 
upward radiation from the disc, accelerating dusty gas up into the stronger lamp-post irradiation until it 
evaporates the dust. This mechanism may well produce a steep rim at the dust sublimation radius, just where we need it to 
reproduce the continuum lags in NGC~5548.

 \begin{figure*}
\textbf{ \Large Irradiating a disc with sinusoidal ripples}
\centering
	\begin{tabular}{@{}cccccc@{}}
 \multicolumn{1}{l}{}\\
	\includegraphics[width=0.47\textwidth]{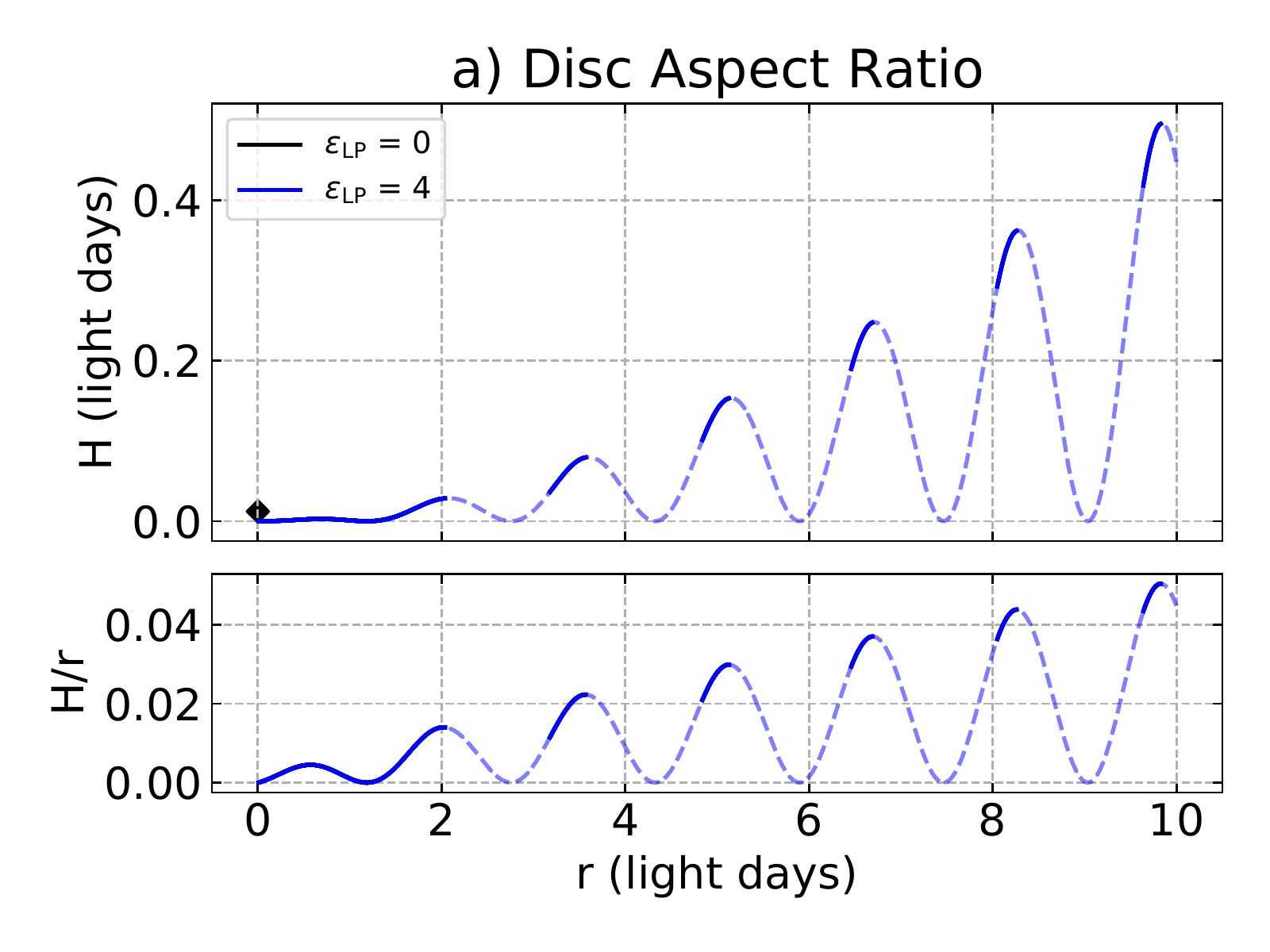}
	\includegraphics[width=0.47\textwidth]{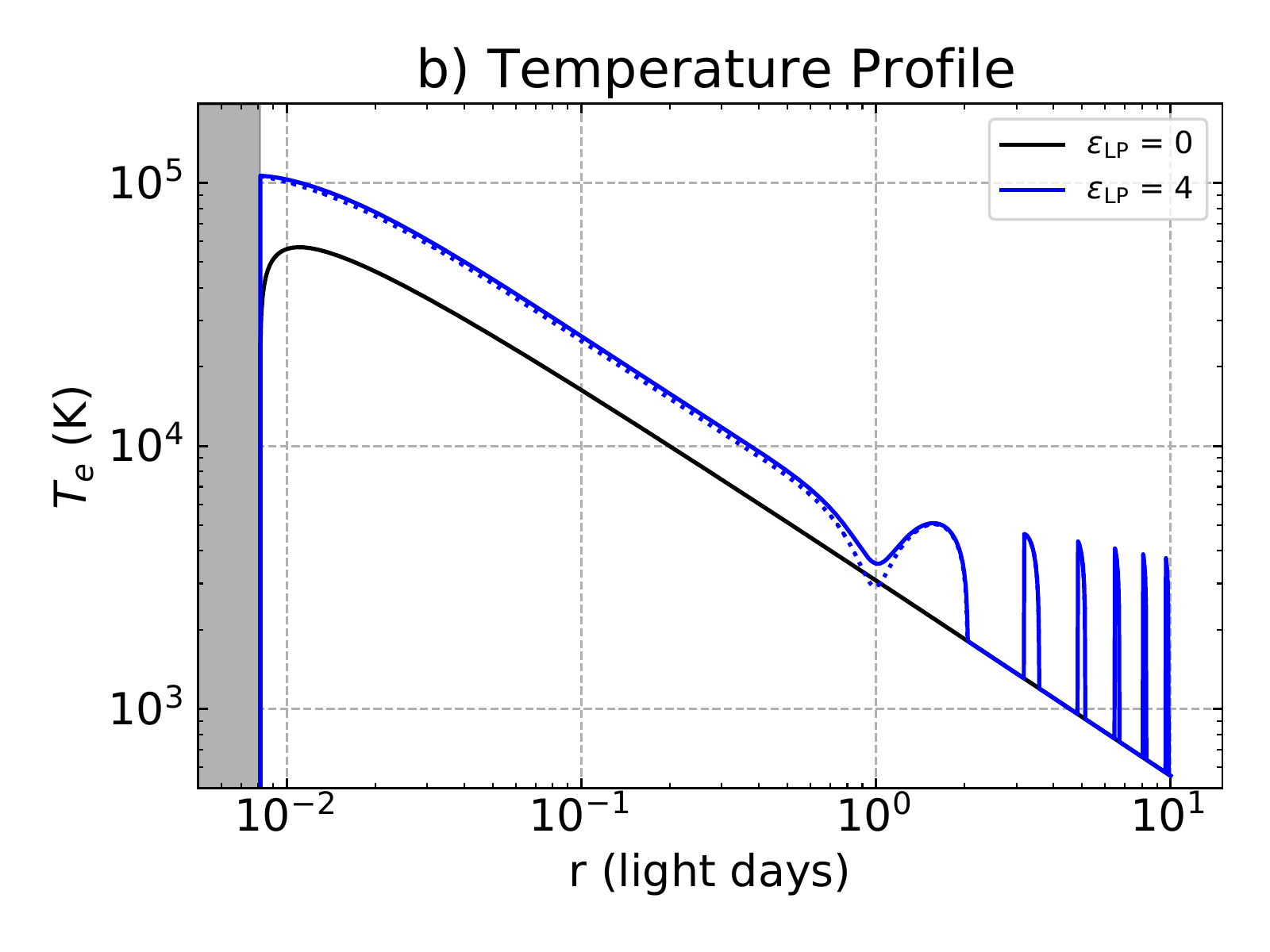}
 \\
  \multicolumn{1}{l}{}\\
	\includegraphics[width=0.47\textwidth]{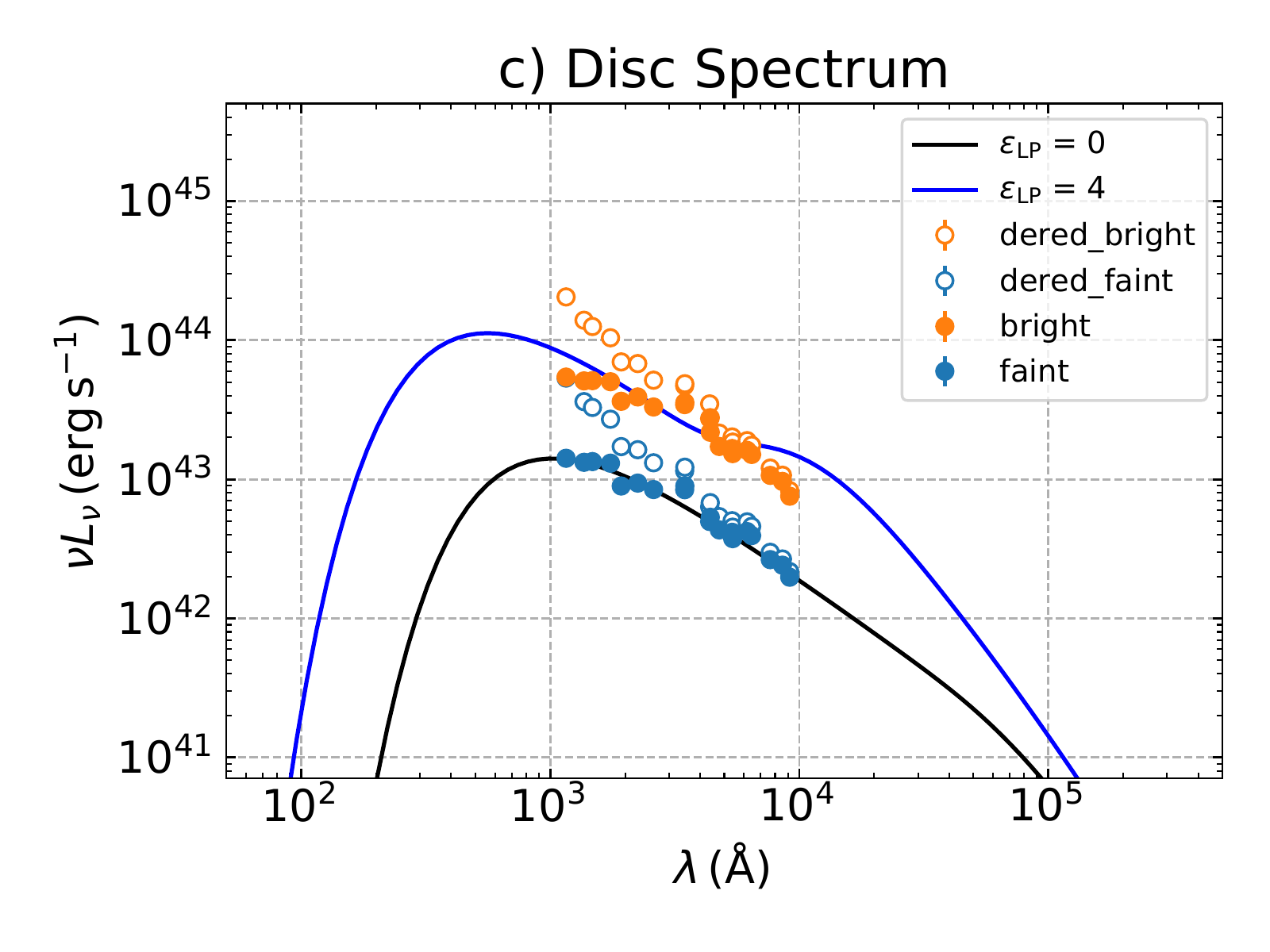}
	\includegraphics[width=0.47\textwidth]{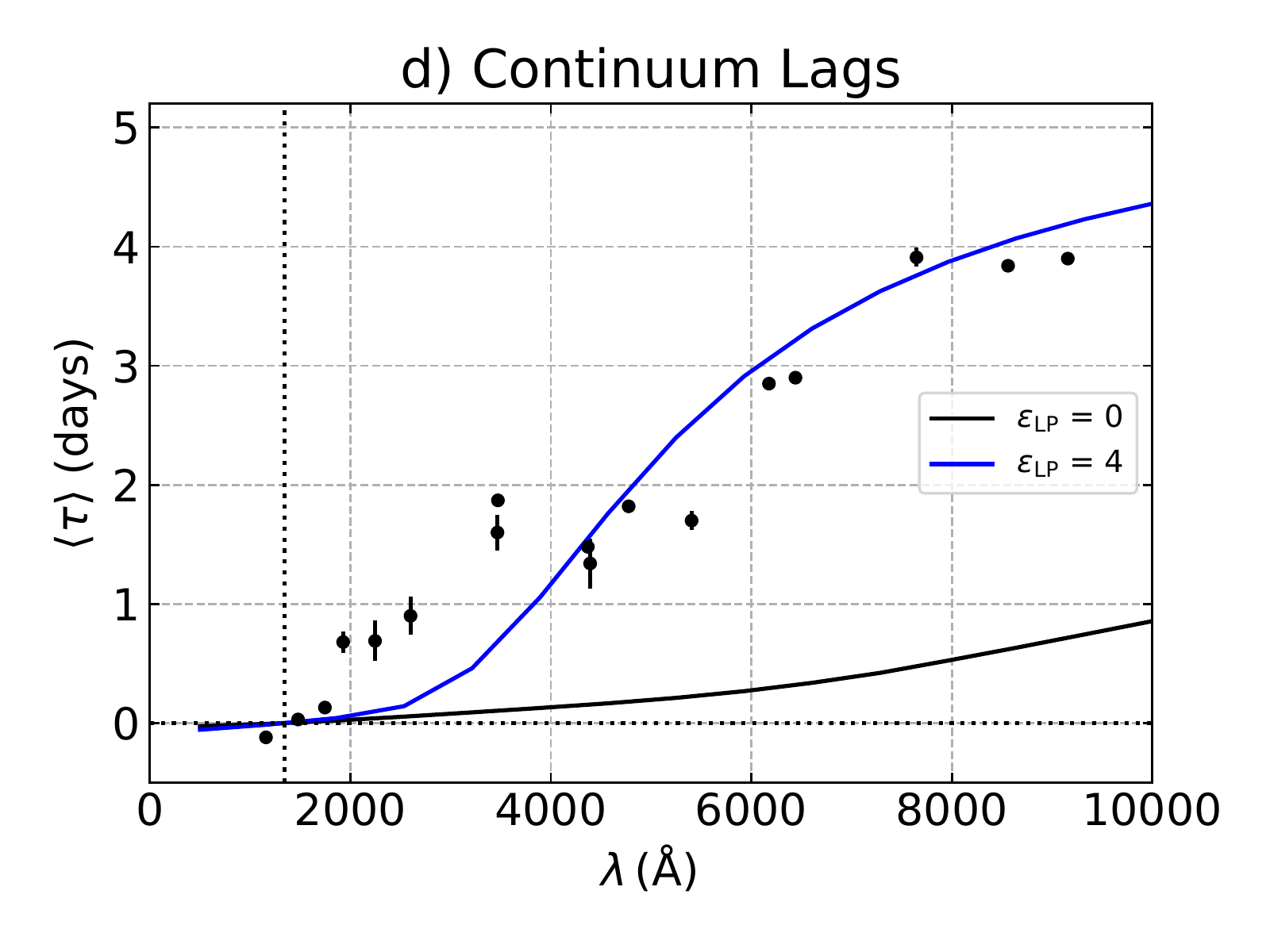}
	\end{tabular}
    \caption{Similar to Figs.~\ref{fig:flat_disc}
    -- \ref{fig:steep} but for a flat disc with sinusoidal wave crests
    (Panel~a).
    The radial wave number $k=4$ radians per light day produces 7 crests inside $10$~light days.
    The wave crest heights scale as $r^{1.8}$, 
    reaching $0.5$ light days at the 7$^{\rm th}$ crest near $r=9.3$~light days.
    The lamp-post height $\Hx=3\,r_{\bullet}$
    allows it to irradiate and heat the inward faces near the top of each crest, leaving most of the disc in shadow where viscous heating
    alone powers the temperature profile (Panel~b).
    With the lamp-post off (black curves), viscous heating gives a model disc spectrum (Panel~c) matching the faint-disc data (blue circles).
    Turning the lamp-post on ($\epslp=4$)
    elevates the 
    disc spectrum to roughly match the bright-disc data (orange circles),
    and greatly increases the model lags (Panel~d).
    This model matches the lag data fairly well, but is somewhat redder than the bright-disc SED data, and has a rather large aspect ratio $H/r=0.05$ at the wave crests.}
     \label{fig:thick_disc_ripple}
\end{figure*}

\section{ A rippled disc surface }
\label{sec:ripples}

Another promising option for achieving higher temperatures over a small fraction of the disc surface, and thus longer lags without excessive flux, is to introduce a rippled structure on the reprocessing surface.
The lamp-post then irradiates the inward-facing
crests, leaving the rest of the disc surface
in shadow. 
This landscape concentrates the irradiation into a smaller surface area, increasing the temperature and thus
the lag, while leaving the valleys at the viscous temperature, keeping the flux low.
A rippled surface thus offers the potential to address
both problems, aiming for a successful trade-off between the lags being too small and fluxes being too large.

We therefore consider the possibility of small-scale ripples
on the power-law $H(r)$ profile of the accretion-disc surface,
analogous to the wake and towering features observed by the Cassini spacecraft on Saturn's rings during its equinox  
\citep{cuzzi2010, cuzzi2018, spitale2010}. 
The possibility of small-scale structures may arise from turbulence, embedded stars \citep{artymowicz1993, good03}, or 
quasi-periodic oscillations \citep{ingram2019}, or corrugated waves \citep{shu1983}, standing or propagating, that modulate the 
height of the reprocessing surface \citep{papaloizou1998, lubow2002}. Such structures may also be spontaneously excited or warped through 
the Lense–Thirring precession induced by the misalignment of a rapidly spinning SMBH's rotation and the disc's angular-momentum vectors
\citep{bardeen1975, papaloizou1983, pugliese2015}, 
tidally perturbed by some embedded companions \citep{artymowicz1993, martin2014},
\changes{
or episodic infall of gas with diverse angular momentum orientation
\citep{schawinski2015}. There are observed properties of AGN discs that have been interpreted as evidence
of warped structures \citep{lawrence2010}.
}
The Lense-Thirring torque is 
linearly proportional to SMBH's spin parameter $a_\bullet$ and the warping radius is an increasing function
of $a_\bullet$.  
With $a_\bullet \sim 0.93$ for NGC 5548 (Table \ref{tab:faintdisc}),  this effect is likely to be
relevant. 
\changes{
There is also observational evidence of warps and other substructures in discs
around young stars \citep{matter2014} and interacting binary stars \citep{fragner2010}.
}

In principle, the wave form depends on poorly constrained viscosity in both the radial and normal
direction of the  disc \citep{pringle1992, papaloizou1995, ogilvie1999, nixon2012,  tremaine2014}.
For modeling convenience, we adopt an idealized prescription
by modulating our power-law disc profile with a wave-like perturbation, such that
the vertical profile becomes
\begin{equation}
	H(r) = H_1 (r)\, \left[ \, 1 + A(r) \, \cos{ (k\, r)} \,\right]
	\ ,
\label{eq:hwave}
\end{equation}
\noindent where $H_1(r)$ is the power-law disc height introduced in Eqn.~(\ref{eq_hpowerlaw}), 
$k$ is the radial wave number, 
$A (r) = A_{\rm w} (r/r_1)^{-\beta_{\rm w}}$ is the fractional wave amplitude
at radius $r$, $\beta_{\rm w}$ is the power index for its radial distribution, and $A(r_1)= A_{\rm w}$.  
We also model non-sinusoidal waveforms
by
replacing
\begin{equation} \label{eqn:cosp}
 \cos{(k\,r)} \rightarrow 
    2 \, \left(  \frac{ 1 + \cos{(k\, r)} } { 2 } 
    \right)^p - 1\ ,
\end{equation}
so that $p>1$ steepens and $0<p<1$ flattens the wave crest.

{\doug{In principle, the perturbation in the disc structure by a population of $N$ 
embedded stars or stellar mass black holes may introduce multiple locally thickened 
arcs.  Gravitational instability in discs with $H/r < < 1$ may also lead to the 
arcs with enhanced density, pressure, and aspect ratio. Under the assumption that
multiple arcs have a uniform distribution with high filling factor in the azimuth 
direction,  we have neglected any departure from axial symmetry in $H(r)$.}}

\begin{figure*}
\textbf{ \Large Irradiating a power-law accretion disc with steep wave crests}
\centering
	\begin{tabular}{@{}cccccc@{}}
 \multicolumn{1}{l}{}\\
	\includegraphics[width=0.47\textwidth]{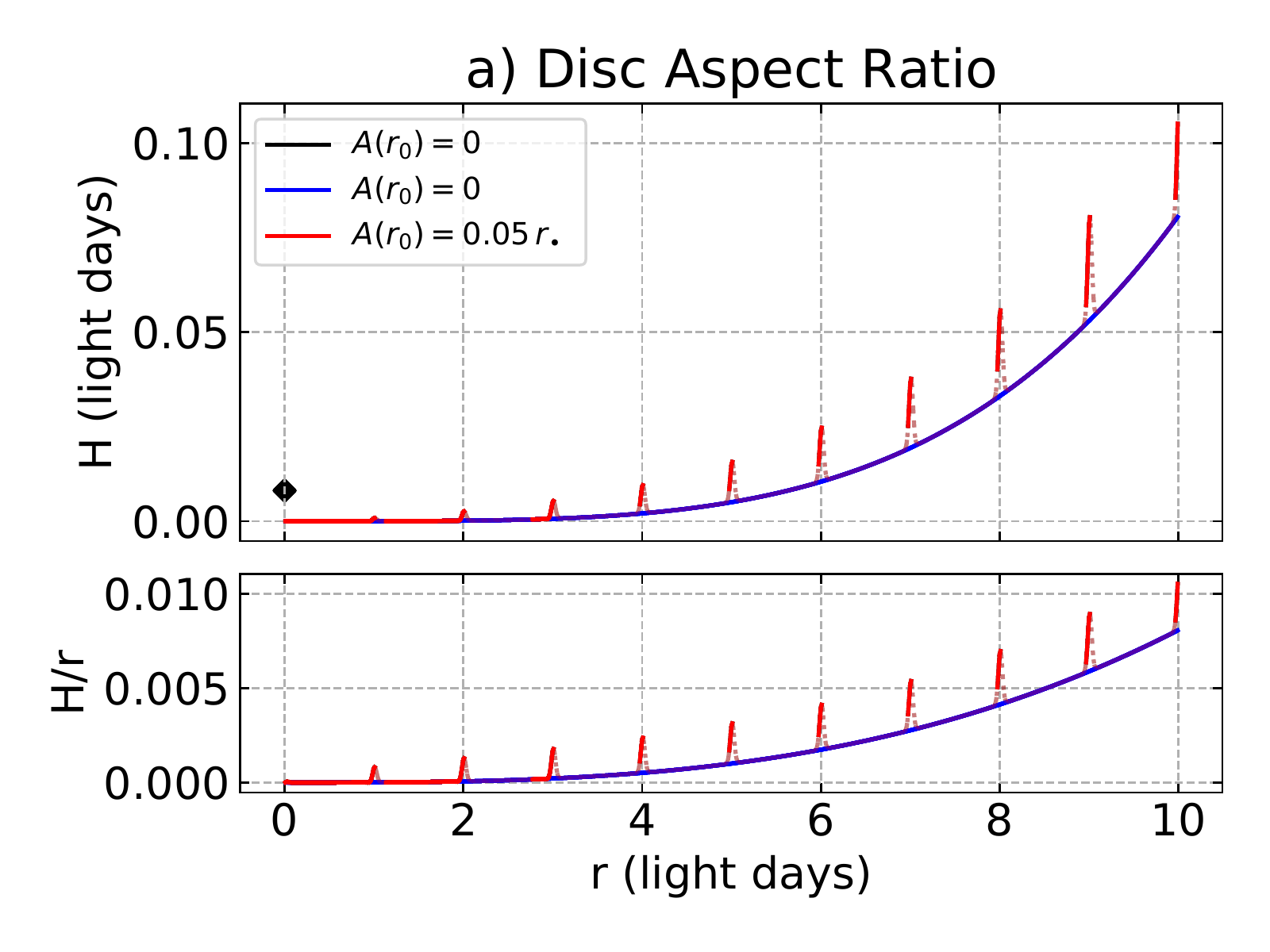}
	\includegraphics[width=0.47\textwidth]{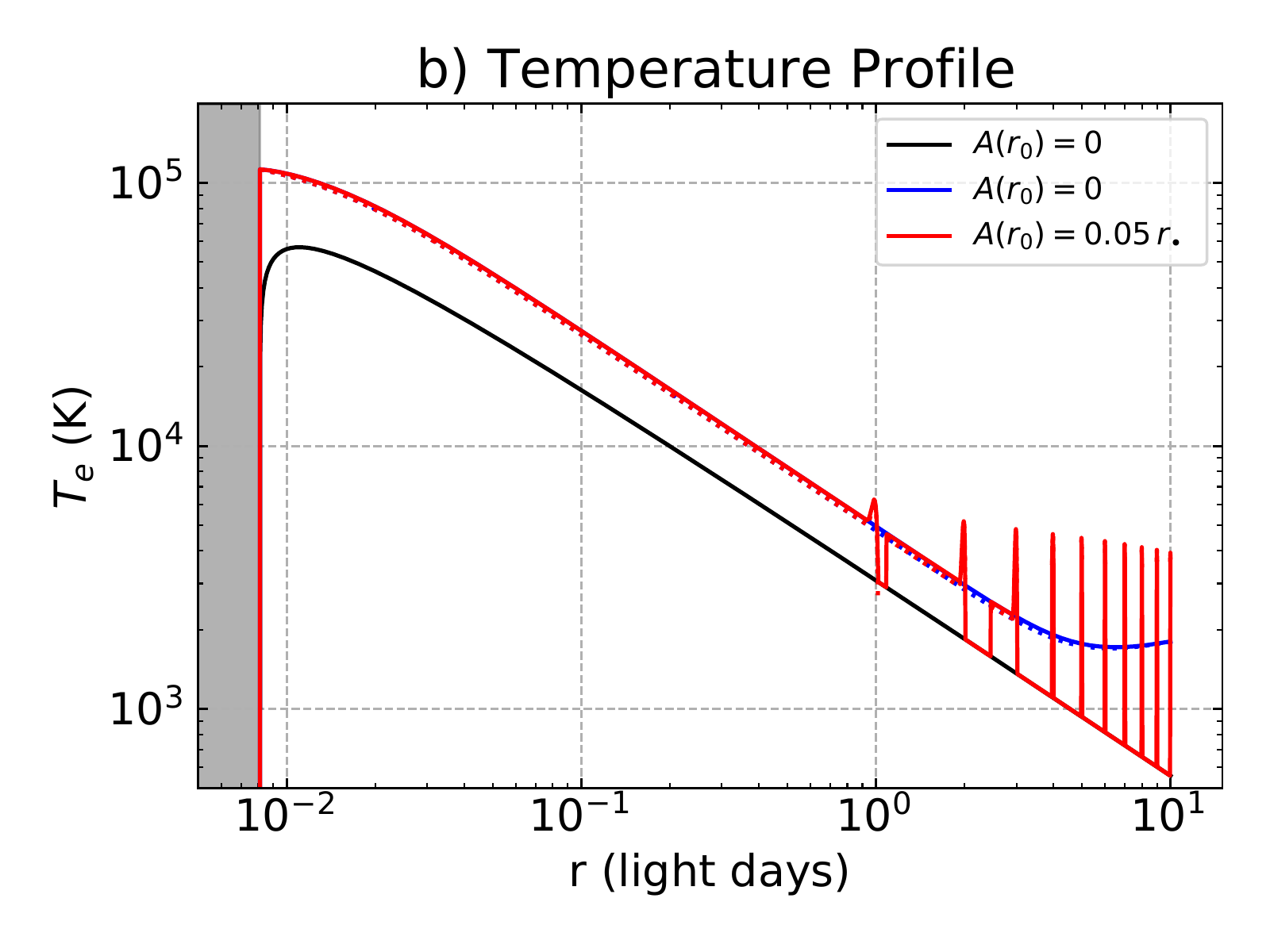}
 \\
  \multicolumn{1}{l}{}\\
	\includegraphics[width=0.47\textwidth]{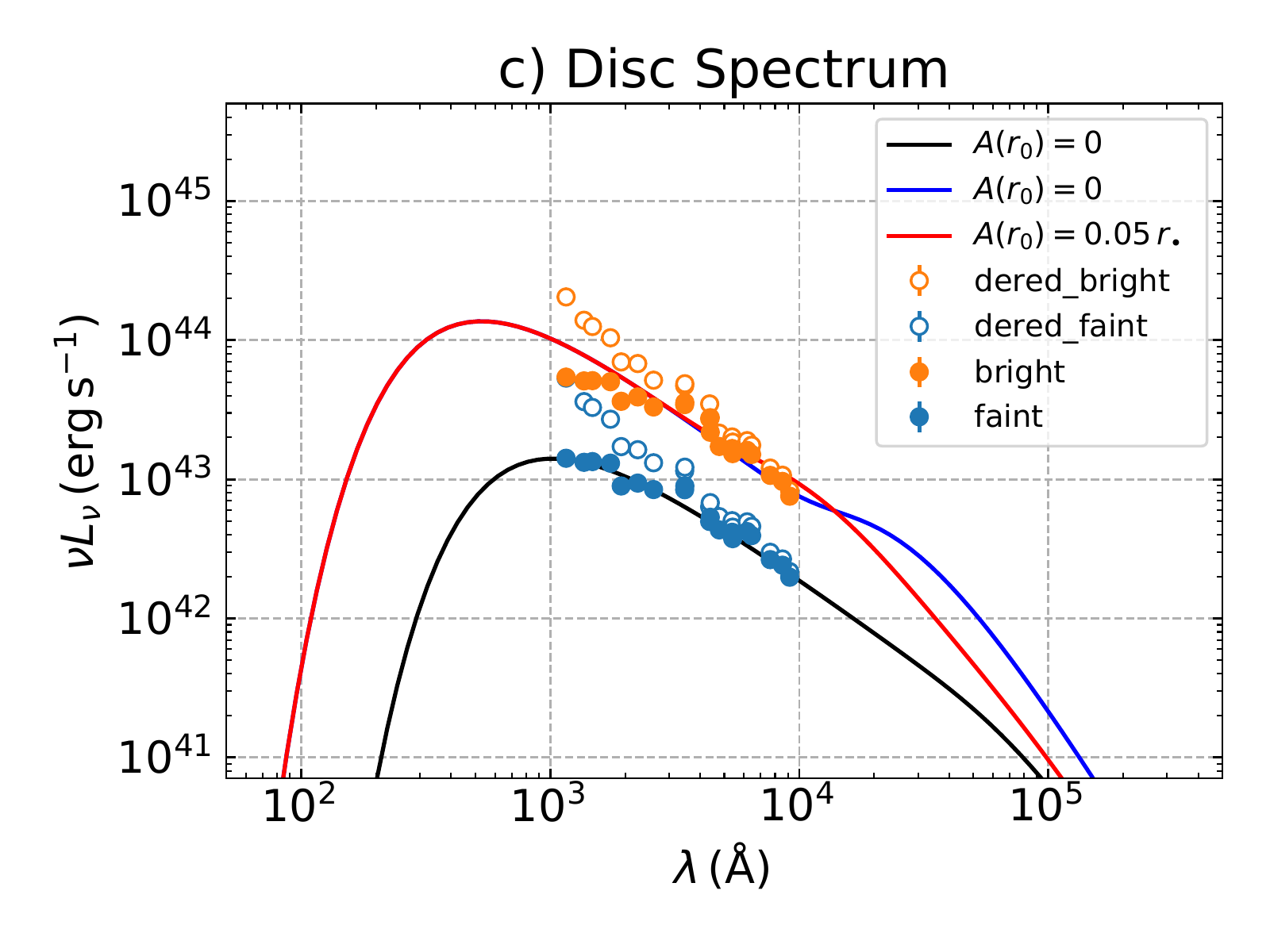}
	\includegraphics[width=0.47\textwidth]{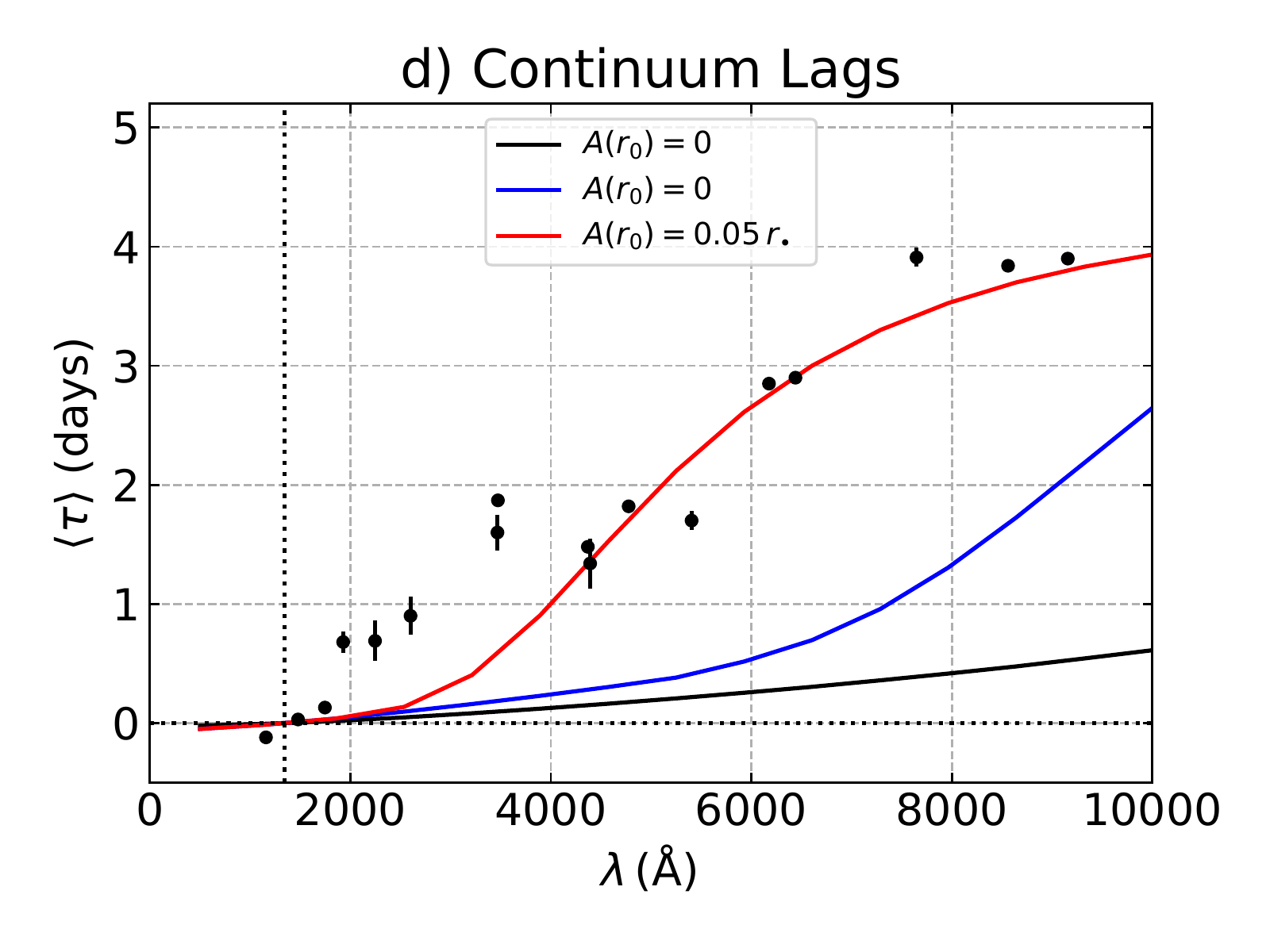}
	\end{tabular}
    \caption{Similar to Figs.~\ref{fig:flat_disc} -- \ref{fig:thick_disc_ripple} but comparing a power-law disc model (blue) with the same model
    plus waves with very sharp crests (red).
    The crests occur at multiples of 1 light day,
    with 10 crests out to 10 light days (Panel~a). Steep crests are produced with
    a power $p = 100$ in Eqn.~(\ref{eqn:cosp}).
    The lamp-post height is $\Hx=3\,r_{\bullet}$.
    With the lamp-post off, viscous heating powers
    a $T_{\rm e}\propto r^{-3/4}$ profile
    that falls to 1000~K near $r_\beta\approx5$~light days
    (black curve in Panel~b).
    Turning the lamp-post on ($\epslp=5$)
    elevates the power-law disc temperatures (blue curve) by a factor of $\sim1.5$ inside $r_\beta$,
    outside which it levels out at $\sim2000~K$. This produces a bump in the model disc SED
    near $\sim2000$~\AA\ (Panel~c)
    and increases the model lags to $3$~days at $10{^4}$~\AA\ (Panel~d).
    With steep wave crests atop the power-law
    (red curves), the irradiated inward faces of the sharp wave crests are heated to $\sim6000$~K, helpfully increasing the model disc lags, but most of the disc surface is in shadow, dropping down to the temperature relation for viscous heating alone
    and decreasing flux on the red end of the SED.
    This model provides a satisfactory fit to the continuum lags and the bright and faint disc SEDs.
    }
    \label{fig:steep_underline}
\end{figure*}

\subsection{Shadow zones between irradiated wave crests}

Consider first the thin-disc limit with 
$\beta_{\rm w}=0$, $H_1(r) = H_1$ constant and small such 
that the disc surface remains below the
lamp-post height, $H_1\, (1 + A_{\rm w}) < \Hx$. 
For large $k\,r$, the $n^{th}$ wave crest occurs at radius
$r_n  = 2\, \pi\, n/k$,
and the change in radius between successive crests is 
$\Delta r = 2\, \pi/k$.
The crests all have the same height $H_c=H_1\,\left(1+A_{\rm w} \right)$.
As viewed from the lamp-post,
the corresponding change $\Delta\theta$ in the small angle 
$\theta=(H_{\rm c}-\Hx)/r$ between 
successive wave crests is
\begin{equation}
	\Delta\theta 
	= \left( \frac{ \Hx - H_1 \, (1 +A_{\rm w}) } { r } \right) \, \frac{ \Delta r }{ r }
	\ .
\end{equation}

\noindent Since $f(r)=r\,\Delta\theta/\Delta r=(\Hx-H_1\,(1+A_{\rm w}))/r$, find

\begin{equation}
T_{\rm e}^4(r)= \left( \frac{ G \, M \, {\dot M} }{  2\, \pi\, \sigma\, r^3} \right)
\left[ \,
\frac{3}{4} +
\epslp  
\left( \frac{ \Hx - H_1\, (1 +A_{\rm w}) }{r_\bullet} \right) \,
\right] \ .
\label{eq:F3}
\end{equation}
Note that $\Delta\theta>0$  for $\Hx > H_1\, (1+A_{\rm w})$. In this limit, all peaks are exposed to the irradiation although the grazing angle $\theta$ between the lamp-post and the plane of the disc surface diminishes with increasing $r$. 

Also in this model, valleys may be in the shadows. Due to reduction in the grazing angle at the elevated wave crests, $f(r)$ is reduced by  $H_1\, A_{\rm w}/r$. 
Within a distance $\delta r$ just inside each crest, the slope leading up to the crest is 
exposed to the lamp-post irradiation if 
\begin{equation}
	\frac{ H_1 \, A_{\rm w} }{  r }\, \left[ \, 1 - \cos{ (k \, \delta r)}\, \right] 
	< \Delta\theta
	\ .
\label{eq:hpeak}
\end{equation}

\noindent After some algebra, the above inequality reduces to
\begin{equation}
\label{eq:drdr}
\frac{ \delta r }{ \Delta r} \approx 
\left( \frac{ \Hx - H_1 \, ( 1 + A_{\rm w} )
	}{  H_1 \, A_{\rm w} \, \pi \, k \, r} \right)^{1/2}
\end{equation}

\noindent in the limit of $k \,\delta r \ll 1$. 
This ratio gives the fraction of the disc surface area that is irradiated.
For a moderately large-amplitude wave, $H_1 = \Hx/2$ and $A_{\rm w} = 1/2$ ,

\begin{equation}
	\frac{\delta r }{ \Delta r} 
	\approx \left( \frac{1 }{ \pi \, k \, r} \right) ^{1/2} 
	= \left( \frac{\Delta r }{ 2 \pi ^2 \, r} \right) ^{1/2}  
\ .
\label{eq:deltar}
\end{equation}
In the small-amplitude limit, $A_{\rm w} \ll 1$, $\delta r$ attains the maximum value $\Delta r$,
as the shadowed area decreases and disappears.
This limit is nearly equivalent to a disc with a ripple-free surface. 
The region between $r_n$ and $r_n+\Delta r - \delta r$ (or equivalently between $r_n$ and $r_{n+1} - \delta r$) is in the shadows so that only the viscous dissipation contributes to the heating.
In this region, $T_{\rm e}=T_{\rm v}$
and the luminosity emitted into $2\,\pi$ sterradians from the shadowed part of the annulus is
\begin{equation}
	\Delta L_{\rm shadow} 
	=\sigma\, T_{\rm v}^4\, 2\, \pi\, r\, \left(\Delta r - \delta r \right)
	= \frac{3 \, G \, M_\bullet \, {\dot M} 
	}{ 4 \, r^2}  \,  \, (\Delta r- \delta r)
	\ .
\end{equation}
In the irradiated region, between $r_{n+1} - \delta r$ and $r_{n+1}$, 
the luminosity emitted is
\begin{equation}
\Delta L_{\rm peak}
	= \frac{G \,M_\bullet {\dot M} }{ r^2} 
	\left( \, 
	\frac{ 3 } { 4 } \, \delta r
	+ \epslp\,
	\frac{
	 \Hx - H_1\, ( 1+A_{\rm w}) }{ r_\bullet }
	\,  \Delta r 
	\right)
\end{equation}

\noindent so that the total luminosity,
$\Delta L_{\rm tot }= \Delta L_{\rm shadow} + \Delta L_{\rm peak}$, is
\begin{equation}
\label{eq_wavyLtot}
	\Delta L_{\rm tot}
	= \frac{ G \, M_\bullet \, \dot{M} }{ r^2} \,
	\left( \frac{3 }{ 4} 
	+ \epslp \, 
	\frac{ \Hx - H_1\, ( 1+A_{\rm w}) }{ r_\bullet } 
	\right)\, \Delta r
	\ . 
\end{equation}

\noindent Thus the presence or otherwise of shadows does not
alter the reprocessed energy, averaged over the radial wavelength, 
other than by the wave amplitude increasing the effective thickness of the disc. 
However, when shadows are present the re-radiation is confined to a smaller surface
area and this results in higher temperatures in the exposed regions
just inside the peaks. 
In thermal equilibrium,
$	\sigma\, T_{\rm peak}^4 \, 2 \, \pi\, r \, \delta r = \Delta L_{\rm peak}$, 
so that 

\begin{equation}
	T_{\rm peak}^4 
	= \left( \frac{ G \, M_\bullet \, {\dot M} } { 2 \pi \sigma \, r^3} \right) 
	\left( \frac{ 3 } { 4 }
	+ \epslp\, \frac{ \Hx - H_1 \, (1+A_{\rm w}) }{ r_\bullet} 
	\, \frac{\Delta r }{ \delta r }
	 \right)
 	\ .
\label{eq:tewave}
\end{equation}

\noindent In the limit of a zero-thickness disc ($H_1=0$, $\delta r = \Delta r$), 
the expression in Eqn.~(\ref{eq:tewave}) reduces to that of Eqn.~(\ref{eq_trprof}). 
More generally (i.e.~for non vanishing $H_1$ and modest $A_{\rm w}$), 
since $\delta r < \Delta r$ (Eqn.~\ref{eq:deltar}), 
the exposed crests are hotter (Eqn.~\ref{eq:tewave})
than in the shadowed valleys (Eqn.~\ref{eq_trprof} with $\epslp =0$).
It is also hotter than that for an irradiated disc with no ripples.

  \subsection{Testing rippled disc models against the data}

Fig.~\ref{fig:thick_disc_ripple} investigates a model with sinusoidal waves atop a flat disc.
This model fits the continuum lags well but has excess flux on the red end of the bright-disc SED.
The wave amplitude here is rather extreme, but note that the valleys in shadow can be ``filled in'' 
without altering the model lags or SEDs, since these shadowed regions remain at the viscous temperature.
The fractional wave amplitude can thereby be reduced without spoiling the success in matching the data.

Fig.~\ref{fig:steep_underline} considers
a rippled-disc model with steeper wave crests atop a concave power-law disc. 
This model fits well the faint and bright disc SEDs and the optical lags but less well the UV lags.
Both of these rippled-disc models achieve 
reasonably good fits to both the NGC~5548 continuum lags and the faint-disc and bright-disc SEDs.
The key to success in fitting the lags
is to elevate disc temperatures to about $6000$~K
at a radius of about 5 light days, but only
on a small fraction of the disc surface, thus
to preventing an unwanted rise on the red end of the SED.

These rippled-disc models have a lot of flexibilty that could be fine-tuned to refine the fits. However, such an analysis is beyond the scope of this paper and may in any case be of limited value at this stage
given the simplifying assumptions and the wide range of physical mechanisms that may lead to waves or warps on accretion disc surfaces.
Our intention here is to illustrate with a few examples how a rippled disc surface can alter the model lags in ways that ease the tension between disc models and the larger-than-expected disc sizes inferred from reverberation mapping time lags.

\section{Summary and Outlook}
\label{sec:conc}
We have introduced a simple analytic model to investigate the effect of curved vertical structure, including ripples and rims, in an AGN accretion disc. The rippled disc structure shadows parts of the accretion disc from irradiation by the hot corona around the black hole. We explore how this shadowing affects the disc SED and show how such a structure is expected to alter the inter-band continuum lags measured from reverberation mapping experiments. 

We test predictions of this model against the disc sizes inferred from light travel time lags from the STORM reverberation mapping study of NGC~5548.
We find models that succeed in fitting not just the disc lags but also the faint-disc and bright-disc SEDs, by introducing 
steep 
transitional or wavy structures on the disc surface.
The irradiation is then concentrated on the relatively small center-facing slopes near a rim or multiple crests, 
increasing the local temperature.
Longer lags then result first due to the increased temperature but also from the front-to-back asymmetry when the inward-tilted crests are viewed at
a finite inclination angle.
Lags increase because
foreshortening reduces our view of the 
irradiated faces on the near side, and 
opens up our view of the same structures on the far side of the disc.
The flux does not increase excessively
on the red end of the SED because 
the rim and/or ripples expose
only a small fraction of the disc surface 
to irradiation, leaving
most of it
close to $T_{\rm e}\propto r^{-3/4}$.

Our approach goes beyond previous work that has focused on fitting the lags without also fitting the disc SED.
The disc fluxes are derived from the spectral variations, extrapolating to lower disc fluxes to correct for the host galaxy contribution.
The disc SED provides an equally important constraint on the accretion disc temperature profile, and is actually easier to measure than the inter-band continuum lags.
A viable interpretation of the observed continuum lags and spectral variations requires reverberating disc models that fit both constraints.

Perhaps the simplest successful model
(Fig.~\ref{fig:steep}) retains a thin disc with viscous heating alone in the faint disc, augmented by lamp-post irradiation in the bright disc, and with a steep rim on which the irradiation elevates the temperature to $\sim6000$~K.
Intriguingly,
this model's intrinsic disc temperature (due to viscous dissipation alone) just inside this steep rim is $\sim1000$~K, rising to
$\sim1500$~K with the additional grazing-incidence lamp-post irradiation.
As these are close to
dust sublimation temperatures, the disc should thicken outside this radius due to vertical 
radiation pressure acting on dust grains in the disc atmosphere. This ``failed disc wind'' 
hypothesis \citep{czerny2011, baskin2018, naddaf2021} can be tested by applying the thin-disc 
+ steep-rim model to the lags and SEDs of other AGN to see if they consistently place the rim near the dust sublimation radius.

\changes{
The failed disc wind model offers a natural mechanism for generating BLR gas clouds outside the radius where our model places the steep disc rim.
Scattering of the disc emission by the BLR inter-cloud 
medium may then introduce additional time delays that may be interpreted with a relatively large lamppost 
height \citep{jaiswal2022}. 

Our blackbody reprocessing model cannot produce the observed excess lag in the $u$~band, which samples the Balmer continuum. To account for this effect, the BLR clouds offer a natural venue where reprocessing yields a diffuse bound-free continuum response with longer lags than those from the more compact accretion disc \citep{Korista2001,Lawther2018,Chelouche2019,Netzer2022}.
Which reprocessor -- the disc or the BLR -- dominates the lags?
\cite{Netzer2022} proposes that BLR lags dominate. Adopting a negligible disc lag, he models AGN lag spectra, with good success, by diluting BLR lags from photo-ionisation models of radiation-pressure confined gas clouds, with a wavelength-dependent mix reflecting the BLR/disc flux ratio.
Our models represent the opposite extreme, with blackbody reprocessing on a structured disc surface dominating the lag spectrum, plus a minor bound-free response from BLR gas clouds to increase lags in the $u$ band. Further intensive multi-band monitoring, combined with careful modelling to fit simultaneously the lag spectrum and the changes in the disc SED, should be able to clarify this ambiguity.
}

Other possible mechanisms for generating steep vertical structures include disc warps and propagating waves.
With $a_\bullet=0.93$ inferred from the SED during  the low state of NGC~5548,
the Lense-Thirring torque is expected to be significant in the disc inner region.
The alternative ripple model 
(Figs.~\ref{fig:thick_disc_ripple}, \ref{fig:steep_underline}) is similar to the structure expected for warped discs 
around a rapidly spinning and slightly misaligned SMBH \citep{bardeen1975, papaloizou1983, pringle1992, 
papaloizou1995, ogilvie1999, nixon2012,  tremaine2014}.

For all the models presented here, the total (disc + lamp-post) luminosity during the bright-disc state is assumed to be sub-Eddington.
For our successful models (Figs.~\ref{fig:steep}, \ref{fig:thick_disc_ripple}, \ref{fig:steep_underline})
we find ($ L_{\rm LP} + L_{\rm disc})/L_{\rm Edd} \approx 0.05$. 

Typical values of the disc aspect ratio {\doug {for the exposed disc surface 
at the base of the ripples is $H/r \leq 10^{-3}-10^{-2}$ (Figs. \ref{fig:steep}, \ref{fig:thick_disc_ripple},  \ref{fig:steep_underline}) at $r=4$ light days. 
The corresponding density scale height at the disc midplane $H_d  (= c_s/\Omega$ 
where $c_s$ is the sound speed at the mid-plane) is several times smaller \citep{garaud2007} 
(i.e. $h_{-2} \equiv 10^2 H_{\rm d}/r < 0.5$). 
With an $\alpha$ prescription for disc's viscosity ($\nu=\alpha \,\Omega \,H_{\rm d}$)
the mass transfer rate in a steady state is
\begin{equation}
    {\dot M}_{\rm d} = 3\, \pi\, \Sigma \,\nu =
 3 \,{\sqrt 2} \,\frac{\alpha\, H_{\rm d}^2 } { Q \,r^2}\, M_\bullet \, \Omega = {\dot M}_\bullet
\end{equation}
where $Q= c_s \,\Omega/\pi \,G\, \Sigma$ is the gravitational stability parameter.
With the model parameters in Table~\ref{tab:faintdisc}, we find $Q/\alpha \simeq 80 \,
h_{-2} ^3$.  Marginal gravitational stability, with $Q \simeq 1$ and 
$\alpha \sim 0.1$ \citep{deng2020}, would be maintained \citep{good03} with $h_{-2} \leq 0.5$.}  }
Based on these inferences, we are constructing a self-regulated stellar evolution and pollution in AGN disc 
(SEPAD) model which will be presented elsewhere. 

In the observed wavelength range ($10^{3-4}$\AA), the peak brightness is 5 times that attributed to 
the steady-state 
viscous dissipation.  In our best-fit models, the required lamp-post bolometric luminosity is $20$ times that
emitted by the entire disc during the faint state.  Such large fluctuation amplitudes in X-ray flux are commonly
observed in AGNs, on the timescales of hours \citep{ponti2012}. Moreover, these sources appear to originate from
the corona regions within a few $r_\bullet$ from the SMBH \citep{chartas2009, kara2016} as we have assumed in
our lamp-post model.  The cause of these large fluctuations may be associated 
with nonlinear variations in the accretion rate near the ISCO due to poorly understood disc 
instability \citep{yuan2014} or modulations resulting from the radiative hydrodynamic processes 
 \citep{jiang2014, jiang2019}.  With the large 
inferred $a_\bullet\,(\simeq 0.93)$, it is also possible to extract the SMBH's rotational energy with
magnetic fields in their proximity \citep{blandford1977, AgolKrolik2000, mckinney2005ApJ}.
Whatever the physical mechanisms driving AGN variability, the increasing number and quality of continuum light curves 
\changes{and long-term monitoring of emission-line shape variability \citep{shapovalova2004}
}
will make such variability studies very popular for investigating the gas flow around the inner disc. Identifying where this departs from the thin-disc theory will be an extremely interesting aspect of study for future reverberation mapping experiments.

\section*{Acknowlegements}

We thank the referee for the helpful comments and suggestion made during the submission of this manuscript.
DL thanks E. Kara and P. Ivanov for useful conversation.  
DAS and KH acknowledge support from the UK Science and Technology Facilities Council (STFC) consolidated grant to St.Andrews 
(ST/M001296/1). DL thanks the Carnegie Foundation for a centenary professorship providing support at St~Andrews where
this investigation was initiated.

\section*{Data Availability}

The data analysed in this paper are for NGC~5548 the cross-correlation time lags from \cite{fa16} and 
the faint-disc and bright-disc 
SEDs from \cite{st17}.



\FloatBarrier

\bibliography{ripdisk} 
\bibliographystyle{mn2e}

\label{lastpage}

\end{document}